\documentclass[journal]{IEEEtran}
\usepackage{color}
\usepackage{dsfont}
\usepackage{multicol}
\usepackage{mathrsfs}
\usepackage{amsfonts}
\usepackage{latexsym}
\usepackage{graphicx}
\usepackage{amsmath}
\usepackage{epstopdf}
\usepackage{subfig}
\usepackage{cite}
\usepackage{enumerate}
\usepackage{amssymb}
\usepackage{bm}
\usepackage{booktabs}
\usepackage{array}

\newtheorem{theorem}{\textbf{Theorem}}
\newtheorem{lemma}{\textbf{Lemma}}
\newtheorem{remark}{\textbf{Remark}}
\newtheorem{definition}{\textbf{Definition}}
\newtheorem{assumption}{\textbf{Assumption}}

\newtheorem{example}{\textbf{Example}}

\newcommand*{\QEDB}{\hfill\ensuremath{\square}}

\begin{document}

\twocolumn[{
\centering
\textbf{\color{red}This work has been submitted to the IEEE for possible publication. 
Copyright may be transferred without notice, after which this version may no longer be accessible.}
\vspace{1em}
}]

\title{
\huge{{Unified Performance Control  for Non-Square Nonlinear Systems with Relaxed Controllability}}
}

\author{Bing Zhou, Kai Zhao, Yongduan Song, \emph{Fellow, IEEE}, and Zhen Chen
\thanks{This work was supported in part by the National Key Research and Development Program of China under Grant 2022YFB4701400/4701401, and in part by the National Natural Science Foundation of China under Grant (No.61991400, No.61991403, No. 62250710167, No.61860206008, No.61933012, and No. 62273286).\emph{ (Corresponding author: Yongduan Song.)}

The authors are with the School of Automation, Chongqing University, Chongqing, 400044, China (bbzhou@cqu.edu.cn; zhaokai@cqu.edu.cn; ydsong@cqu.edu.cn; chenzhen211@stu.cqu.edu.cn)
}
}

\markboth{}
{Shell \MakeLowercase{\textit{et al.}}: Bare Demo of IEEEtran.cls for Journals}

\maketitle

\begin{abstract}
In this paper, we investigate the problem of unified prescribed performance tracking for a class of non-square strict-feedback nonlinear systems under relaxed controllability conditions. By using a skillful matrix decomposition and introducing some feasible auxiliary matrices, a more generalized controllability condition than the current state of the art is constructed, which can be applied to both square and non-square nonlinear systems subject to actuator faults and unknown yet time-varying control gain. Incorporating the relaxed controllability conditions and the uniform performance specifications into the backstepping design procedure, a prescribed performance fault-tolerant controller is developed that can achieve different performance demands without modifying the controller structure, which is more flexible and practical.
In addition, the destruction of the system stability by unknown controllability auxiliary matrices and unknown nonlinearities is circumvented by embedding the available core information of the state-dependent uncertainties into the design procedure. Both theoretical analysis and numerical simulation demonstrate the effectiveness and benefits of the proposed method.
\end{abstract}


\IEEEpeerreviewmaketitle

\section{Introduction}
Over the past decade, significant progress has been made in the field of adaptive control for uncertain nonlinear systems, with applications spanning robotics \cite{HeC2021tac}, autonomous vehicles \cite{LiX2023tac1}, quadrotors \cite{TianB2019tie}, and more.
It should be emphasized that the primary challenge prior to control design for systems with uncertain nonlinearities is to establish a suitable controllability condition, that is, to impose appropriate restrictions on the unknown control gains to guide controller design and stability analysis.

For single-input single-output systems, a common controllability condition is to assume, without loss of generality, that there exists a constant lower bound on the control gain $g$. However, such a condition is usually not applicable or conservative for multiple-input multiple-output (MIMO) systems since the control gain is a matrix.
In order to solve the control problems of this type of system, some classical assumptions have been imposed in the existing literature.
To name a few, one classical assumption imposed on the control gain matrix $\bm{g}$ is requiring that $\bm{g}$ is symmetric and positive definite (SPD) \cite{JinX2017rnc}, which is mathematically elegant but only applicable to a few practical systems, such as robotic systems \cite{Lewis1995tnnls}.
Another improved version of the assumption has been proposed in \cite{XuH2003tac}, i.e., assuming that $\bm{g}+\bm{g}^T$ is uniformly positive definite, which generally aligns with the physical models of several practical systems, such as robotic systems \cite{Lewis1995tnnls}, the wheeled inverted pendulum systems \cite{LiZ2010auto}, and thus its validity has been widely recognized \cite{Bechlioulis2008tac,SongY2016tac,Katsoukis2021tac}.

However, these controllability conditions (the restrictive assumptions imposed on the control gain matrix) are not always satisfactory.
On the one hand, not all practical systems satisfy such conditions, e.g., high-speed train systems \cite{SongY2016tie} and quadrotors \cite{TianB2019tie}, since the control gain matrix $\bm{g}$ in their dynamics is neither SPD nor satisfies the fact that $\bm{g}+\bm {g}^T$ is SPD.
On the other hand, it is known that actuator failure usually destroys this controllability condition.
Specifically, when actuators experience multiplicative failures, the original $\bm{g}$ will be right-multiplied by an actuation effectiveness matrix $\bm{\rho}$, as mentioned in \cite{ZhangJ2020auto}, the strong controllability condition established in \cite{XuH2003tac} might be invalid.
A number of noteworthy efforts have been made to establish more relaxed conditions of controllability. For example,  by proposing a sector condition with a design parameter $\bm{K}$, i.e., assuming that $\bm{K}\bm{g}+\bm {g}^T\bm{K}^T$ is uniformly positive-definite, the control problem of quadrotors can be solved \cite{Lee2016scl}.
To account for the destruction of controllability by actuator failures, some works have embedded the actuation effectiveness matrix $\bm{\rho}$ into the controllability condition in \cite{XuH2003tac}, i.e., assuming that $\bm {g}\bm{\rho}+\bm{\rho}\bm {g}^T$ is uniformly positive definite \cite{JinX2019tcyb}, but it is extremely demanding on the parametric properties of $\bm {g}$ and $\bm{\rho}$, making it hard to satisfy in practice.
Alternative approaches have incorporated the actuation effectiveness matrix into the Lyapunov function, requiring the actuation effectiveness matrix to be differentiable with respect to time \cite{ZhangJ2020auto,ZhaoK2023auto}. 

Inspired by the above discussion, in this paper, we focus on solving the problem of prescribed performance control (PPC) of strcit-feedback MIMO systems under a more relaxed controllability condition. Unlike the existing literatures on PPC for MIMO systems (e.g., \cite{Bechlioulis2008tac,Katsoukis2021tac,JiR2021tcy}), the system considered may have a non-square control gain matrix, and there may be faulty actuators that violate the controllability assumption imposed in these literature.
The core technique of this paper is to decompose the control gain matrix and construct several auxiliary (not necessarily known) matrices to reconstruct the controllability of the system in the control design. The considered problem is well addressed by developing a framework for compatibility of this technique with PPC and backstepping. It is worth noting that the proposed approach differs from classical PPC methods in that it does not rely on a closed-loop initialization set, nor does it require control redesign or reanalysis when performance requirements change.
Preliminary result of controllability relaxation for square nonlinear systems is presented in \cite{ZhouB2023tac}, where the classical PPC is considered. In this paper, we further relax the controllability condition for non-square systems subject to actuator faults, while eliminating the restrictions of the classical PPC so that prescribed performance tracking control with a unified framework can be achieved.
The main features and contributions of this work include:
\begin{enumerate}[\indent (1)]
\item We construct a more general controllability condition than those in existing approaches \cite{ZhouB2023tac,ZhouB2024rnc,LiX2023tac1}, enabling the proposed method to accommodate a broader class of system models, including both square and non-square cases with unknown, time-varying control gain matrices, even in the presence of intermittent actuator faults.
\item The proposed PPC method supports more general and practical performance specifications without conflicting with the relaxed controllability conditions. Unlike existing methods \cite{Bechlioulis2008tac,ZhangJ2020auto,ZhangJ2020smc,LiL2024smc}, which typically handle only symmetric performance functions, our approach enables both symmetric and asymmetric behaviors through simple tuning of design parameters, without modifying the control structure or reanalyzing stability.
\item In contrast to previous works \cite{ZhangJ2020auto,ZhaoK2023auto}, we embed the auxiliary matrix, rather than the actuation effectiveness matrix, into the Lyapunov function. By exploiting the state-dependent structure in the backstepping design, this approach overcomes stability challenges arising from the generalized controllability conditions while ensuring boundedness of all closed-loop signals.
\end{enumerate}

\emph{Notation:} Boldface letters denote matrices or vectors. For a nonsingular matrix $\bm{\mathcal{M}} \in \mathbb{R}^{m \times m}$,
$\underline{\sigma}(\bm{\mathcal{M}})$ and $\lambda_{\min}(\bm{\mathcal{M}})$ represent its minimum singular value and minimum eigenvalue, respectively.
$\bm{I}_m \in \mathbb{R}^{m \times m}$ denotes the identity matrix, and $\bm{0}_m \in \mathbb{R}^m$ is the zero vector.
The symbol $|\cdot|$ indicates the standard Euclidean norm. Finally, $\bm{\Omega}\ast \subset \mathbb{R}^n$ denotes a compact set.

\section{Problem Statement}\label{s24}
Consider a class of high-order MIMO uncertain nonlinear systems in strict-feedback form
\begin{flalign}
\dot{\bm{x}}_{k}=\ &\bm{f}_k(\bm{X}_k)+\bm{g}_k(\bm{X}_k,t)\bm{x}_{k+1}+{\bm{d}_k}(\bm{X}_k),\notag\\
\dot{\bm{x}}_{N}=\ &\bm{f}_N(\bm{X}_N)+\bm{g}_N(\bm{X}_N,t)\bm{u}_a+{\bm{d}_N}(\bm{X}_N)\notag\\
\bm{y}=\ &\bm{x}_1,\ \ k=1,\ldots,N-1\label{os}
\end{flalign}
where $\bm{x}_{i}=[x_{i1},\ldots,x_{in}]^T \in \mathbb{R}^{n}$ ($ i=1,\ldots,N$) is the $i$th state vector;
$\bm{X}_i = [\bm{x}_1^T,\ldots,\bm{x}_i^T]^T \in \mathbb{R}^{n\times i}$;
${\bm{y}}= [y_1, \ldots , y_n]^T \in \mathbb{R}^n$ is the output vector;
${\bm{f}_i}: \mathbb{R}^{in} \rightarrow\mathbb{R}^{n}$ is some nonlinear uncertainty;
${\bm{d}_i}: \mathbb{R}^{in} \rightarrow\mathbb{R}^{n}$ denotes external disturbance; $\bm{u}_a=[u_{a1},\ldots,u_{am}]^T \in \mathbb{R}^m$ is the input vector (the output of actuators); ${\bm{g}_i}: \mathbb{R}^{in}\rightarrow \mathbb{R}^{n \times m}$, $n\leq m$ is an unknown control gain matrix.

As unanticipated actuator faults may occur for ``long-term" operation \cite{TaoG2001tac},  in such case, the abnormal actuator input-output model can be described as
 \begin{equation}
 \label{ua}
 \bm{u}_{a}=\bm{\rho}(t)\bm{u}+\bm{\upsilon}(t)
\end{equation}
where $\bm u=[u_{1},\ldots,u_{m}]^T \in \mathbb{R}^m$ is the designed control input; $\bm{\rho}=\textmd{diag}\{\rho_1,\ldots,\rho_m\} \in \mathbb{R}^{m \times m}$ is the actuation effectiveness matrix that does not need to be piecewise continuous; $\bm{\upsilon}=[\upsilon_1,\ldots,\upsilon_m]^T \in \mathbb{R}^{m}$ represents the bounded time varying bias fault, i.e., $\|\bm{\upsilon}\|<\bar{\upsilon}<\infty$ with $\bar{\upsilon}$ being a positive constant.
In this subsection, we consider the partial loss of effectiveness (PLOE) case, i.e., $\rho_j \in (0,1], j=1,\ldots,m$. Such a scenario may occurs when the attitude control actuators of the aircraft or spacecraft are rotationally limited due to long-term operation \cite{WangX2025smc,ShaoX2023auto}.

Denote the desired trajectory as $\bm{y}^*=[y^*_1,\ldots,y^*_m]^T \in \mathbb{R}^m$, and define the tracking error as $\bm{e}=\bm{x}_1-\bm{y}^*=[e_1,\ldots,e_m]^T$.
The control objective in this paper is to design a control law such that
\begin{enumerate}
  \item All signals in the closed loop systems are bounded in the presence of actuator faults;
  \item The tracking error $\bm{e}$ is always within the prescribed performance bound, i.e.,
      \begin{equation}\label{ppb}
        \mathcal{H}(-\underline{\delta}_j\varphi_j(t))<e_j(t)<\mathcal{H}(\bar{\delta}_j\varphi_j(t)), \ j=1,2,\ldots,m
      \end{equation}
      where $0 < \underline{\delta}_j$, $\bar{\delta}_j \leq 1$ are some positive constants, $\mathcal{H}$ is a time-varying boundary function, and $\varphi_j$ denotes a time-varying scaling function. The detailed definitions of $\mathcal{H}$ and $\varphi_j$ can be found in Section \ref{s3a}.
  \end{enumerate}

To achieve this objective, the following assumptions are needed.
\begin{assumption}\label{A1}
  The desired tracking trajectory $\bm{y}^*$ and its derivatives up to $n$th order are known and bounded.
   The system state vector $\bm{x}_i$ is available for control design.
\end{assumption}
\begin{assumption}\label{A4}
The control gain matrix $\bm{g}_i$ can be decomposed as
\begin{equation}\label{bn}
 \bm{g}_i(\bm{X}_i,t)=\bm{A} \bm{b}_i(\bm{X}_i,t), n<m,\ i=1,\ldots,N
\end{equation}
where $\bm{A}=[\bm{I}_n, \bm{\Lambda}_1,\ldots,\bm{\Lambda}_{(m-n)}]\in\mathbb{R}^{n \times m}$ is a known actuator allocation matrix, with $\bm{\Lambda}_j=[\iota_{j1},\ldots,\iota_{jn}]^T\in \mathbb{R}^{n}$ being a nonzero vector satisfying $\iota_{jp}\geq0$ ($j=1,\ldots,m-n$, $p =1,\ldots,n$), and $\bm{{b}}_{i}\in \mathbb{R}^{m\times m}$  is an extension matrix of $\bm{g}_i$. Without loss of generality, assume ${\bm{g}_k}\in \mathbb{R}^{n \times n}$ ($k =1,\ldots,N-1$) and ${\bm{g}_N}\in \mathbb{R}^{n \times m}$ with $n< m$.
\end{assumption}
\begin{assumption}\label{A2}
{There exists an unknown positive constant $\bar{g}_j$ such that $\|\bm{g}_i\|\leq\bar{g}_i$ ($i=1,\ldots,N$). Further, for $\bm{g}_j$ ($j=1,\ldots,N-1$), there exist an unknown diagonal and positive definite matrix $\bm{P}_1(t) \in \mathbb{R}^{n \times n}$ and some unknown symmetric and positive definite matrices $\bm{P}_k(\bm{X}_{k-1},t) \in \mathbb{R}^{n \times n}$ ($k=2,\ldots,N-1$)
such that $\bm{G}_j:={\bm{P}_j \bm{g}_j+\bm{g}_j^T\bm{P}_j}$ ($j=1,\ldots,N-1$) are uniformly sign-definite with known signs, i.e., $0<\lambda_i \leq{\underline{\sigma}}(\bm{G}_i/2)$,
where $\lambda_i$ is an unknown constant; and for $\bm{g}_N$, there exists an unknown symmetric and positive definite matrix $\bm{P}_{N}(\bm{X}_{N-1},t) \in \mathbb{R}^{n \times n}$
such that $\bm{G}_{N}:={\bm{P}_{N} \bm{{g}}_{N}\bm{\rho}\bm{A}^T+\bm{A}\bm{\rho}\bm{{g}}_{N}^T\bm{P}_{N}}$ is uniformly sign-definite with known sign, i.e., $0<\lambda_{N} \leq{\underline{\sigma}}(\bm{G}_{N}/2)/\|\bm{A}\|$, where $\lambda_{N}$ is an unknown constant. Without loss of generality, we assume that all $\bm{G}_{i}$, $i=1,\ldots,N$, are uniformly positive definite.}
\end{assumption}

\begin{remark}\label{R1}
{\emph{Assumption} \ref{A2} ensures the controllability of system (\ref{os}), which is milder than the current state-of-the-art \cite{XuH2003tac,Lee2016scl,Katsoukis2021tac,MengW2021tcy,HuangX2022tac,ZhouB2023tac}. In \cite{Lee2016scl} and \cite{HuangX2022tac}, similar conditions are considered, yet $\bm{P}_i$ is required to be known and diagonal form, respectively, and the condition in \cite{ZhouB2023tac} is only available for square systems. Thus, the controllability conditions in \cite{XuH2003tac,Lee2016scl,ZhouB2023tac} can be viewed as some special cases of \emph{Assumption} \ref{A2} since $\bm{P}_i$ is allowed to be unknown and can handle non-square nonlinear systems in our case.
Moreover, it is worth noting that the auxiliary matrix $\bm{P}_i$ is only required for stability analysis, as opposed to needing to be present in the control design as in \cite{Lee2016scl}, which thus benefits the implementation of our method.}
\end{remark}
\begin{remark}\label{R2}
The decomposition in \emph{Assumption}~\ref{A4} represents a simplified input-output mapping, where $\bm{I}_n$ denotes the primary input contribution, and $\bm{\Lambda}_j$ captures the effect of additional inputs, such as redundant actuators or extra degrees of freedom. The specific form of $\bm{\Lambda}_j$ in $\bm{A}$ depends on the system's redundant dynamic allocation. For example, $\bm{\Lambda}_j = [1/\sqrt{3},\, 1/\sqrt{3},\, 1/\sqrt{3}]^T$ implies that the redundant input contributes equally to all primary channels, whereas $\bm{\Lambda}_j = [0,\, 0,\, 1]^T$ indicates that it only affects one of them. Such decomposition strategies are commonly employed in redundant control systems, including robotics and spacecraft \cite{ShaoX2022jgcd}. It is worth emphasizing that the adopted decomposing strategy is only valid for over-actuated systems with $n < m$. Under this condition, as long as $\bm{A}$ is full row rank, the decomposition can represent any $\bm{g} \in \mathbb{R}^{n \times m}$ without loss of generality, thereby avoiding modeling errors or structural limitations. In contrast, for the under-actuated case $n > m$, the column space of $\bm{A}$ is restricted to at most an $m$-dimensional subspace of $\mathbb{R}^n$, which limits the admissible range of $\bm{g}$ and prevents the coverage of all possible configurations. This structural limitation undermines both controllability analysis and controller design generality.
Therefore, to avoid such degeneracy and potential singularities in the modeling framework, this work exclusively focuses on non-square MIMO systems with redundant inputs ($n < m$), under which the proposed control strategy and the associated extended controllability conditions are developed.
\end{remark}
\begin{remark}
Unlike \cite{LiX2023tac1} and \cite{SongY2016tac}, where $\bm{g}_N$ is explicitly decomposed, \emph{Assumption} \ref{A4} requires no additional information about $\bm{g}_N$ beyond structural characteristics. That is, $\bm{A}$ only reflects predefined dynamic allocation and does not ensure system controllability.
In \cite{LiX2023tac1} and \cite{SongY2016tac}, the controllability condition is established by decomposing the control gain matrix to obtain a system information-based matrix $\bm{L}\in\mathbb{R}^{n \times m}$ with full row rank, however, such a decomposition is complex and hard to implement for some systems, moreover, such a controllability condition may be invalidated due to the presence of actuator faults (see \emph{Example 2} for detail).
It is worth emphasizing that by \emph{Assumptions} \ref{A4} and \ref{A2}, the controllability of the system is ensured by matching the auxiliary matrix $\bm{P}_i$ analytically, rather than by solving the matrix $\bm{L}$ in a complex offline manner, as seen in our later development.
\end{remark}

To validate the effectiveness of \emph{Assumption}\ref{A4} and \emph{Assumption}\ref{A2}, two examples are presented to show how they facilitate the relaxation of standard controllability assumptions and improve the applicability of the control design.
\begin{example}
(Square case) Consider
 \begin{align*}
  \bm{g}=\begin{bmatrix}
    2 & 1\\
    5 & 3
  \end{bmatrix},\ \textrm{and} \  \bm{g}+\bm{g}^T=\begin{bmatrix}
    4 & 6 \\
    6 & 6
  \end{bmatrix}.
  \end{align*}
 It is easy to confirm that neither $\bm{g}$ nor $\bm{g} + \bm{g}^T$ are SPD, which implies that the conventional controllability assumptions employed in \cite{XuH2003tac,Katsoukis2021tac} do not hold in this case. However, by selecting an appropriate auxiliary matrix 
   \begin{align*}
  \bm{P}=\begin{bmatrix}
    3 & 0\\
    0&1
  \end{bmatrix},\ \textrm{then} \  \bm{P}\bm{g}+\bm{g}^T\bm{P}=\begin{bmatrix}
    12 & 8 \\
    8 & 6
  \end{bmatrix}
  \end{align*}
 which is symmetric positive definite. This shows that \emph{Assumption}~\ref{A2} holds, thus guaranteeing the controllability of the system under the relaxed condition.
\end{example}

\begin{example}
(Non-square case)
 Consider a spacecraft equipped with four reaction wheels (RWs) \cite{ShaoX2023auto}, and the control gain matrix is defined as $\bm{g}=\bm{J}^{-1}\bm{D}\in \mathbb{R}^{3 \times 4}$, where the inertia matrix $\bm{J}\in \mathbb{R}^{3 \times 3}$ and configuration matrix $\bm{D}\in \mathbb{R}^{3 \times 4}$ are given by
  \begin{align*}
  \bm{J}=\begin{bmatrix}
    20 & 1.2 &0.9 \\
    1.2& 5  &1.4\\
    0.9&1.4&5
  \end{bmatrix}\textrm{kg}\cdot \textrm{m}^2,\  \bm{D}=\begin{bmatrix}
    1 & 0 &0&{1/\sqrt{3}} \\
    0 & 1 &0&{1/\sqrt{3}} \\
    0 & 0 &1&{1/\sqrt{3}}
  \end{bmatrix}.
  \end{align*}
 Assume that the actuation effectiveness matrix is set as $\bm{\rho}(t)=\textmd{diag}\{0.1-0.08\sin(t),0.2-0.17\sin(t),0.2-0.15\sin(t),0.2+0.18\sin(t)\}$. If we choose $\bm{A}=\bm{D}$ as in \cite{ShaoX2022jgcd} and \cite{ShaoX2023auto}, by a simple calculation, it can be verified that $\bm{G}^*={\bm{{g}}\bm{\rho}\bm{D}^T+\bm{D}\bm{\rho}\bm{{g}}^T}$ is not positive definite, but we can still make \emph{Assumption} \ref{A2} hold by choosing
 \begin{align*}
  \bm{P}=\begin{bmatrix}
    0.7+0.1\sin(t) & 0 & 0  \\
    0 & 0.1 & 0  \\
    0 & 0 & 0.6+0.1\cos(t)
  \end{bmatrix}.
  \end{align*}
Consequently, the system maintains controllability under actuator degradation and varying input effectiveness, demonstrating the robustness and flexibility of the proposed framework for non-square, over-actuated MIMO systems such as spacecraft with multiple reaction wheels.
\end{example}
\begin{assumption}\label{A3}
  There exist some unknown positive constants $a_{fi}$, $a_{11}$, $a_{12}$, $a_{k1}$, $a_{k2}$ and some known nonnegative scalar ``core functions" $\phi_{fi} (\bm{X}_i)$, $\phi_{k1} (\bm{X}_{k-1})$ and $\phi_{k2} (\bm{X}_k)$ ($i=1,2,\ldots,N$, $k=2,\ldots,N$) such that
 \begin{align}
    &\|\bm{f}_i(\bm{X}_i)+{\bm{d}_i}(\bm{X}_i)\|\leq a_{fi} \phi_{fi}(\bm{X}_i)\label{fphi}\\
    &\|\bm{P}_{1}(t)\|\leq a_{11},\ \big\|\frac{\partial \bm{P}_{1}(t)}{\partial t}\big\|\leq a_{12}\label{m1}\\
    &\|\bm{P}_{k}(\bm{X}_{k-1},t)\|\leq a_{k1}\phi_{k1} (\bm{X}_{k-1}),\notag\\
     &\big\|\frac{\partial \bm{P}_{k}(\bm{X}_{k-1},t)}{\partial t}\big\|\leq a_{k2}\phi_{k2} (\bm{X}_{k})\label{phi2}
  \end{align}
 where $\phi_{fi}$, $\phi_{k1}$ and $\phi_{k2}$ are radially unbounded.
\end{assumption}

\begin{remark}\label{R3}
\emph{Assumption} \ref{A3} is not a stringent requirement, and many practical systems can readily fulfill such a condition by extracting crude model information \cite{SongY2016tac,ZhaoK2023auto,LiX2023tac1}.
Particularly, even if no "core function" can be extracted, one can judiciously choose $\phi_{fi}=\phi_{k1}=\phi_{k2}=1$, which is  equivalent to impose some unknown upper bounds on $\|\bm{f}_i\|$, $\|\bm{d}_i\|$, $\|\bm{P}_{k}\|$ and $\|{\partial \bm{P}_{k}}/{\partial t}\|$. Consequently, by employing core functions, non-parametric uncertainties can be handled elegantly with less restrictions.
\end{remark}

\section{Main Result}
\subsection{Non-monotonic Performance Function}\label{s3a}
To achieve the prescribed tracking performance, we first introduce the rate function $\beta(t)$ satisfying the following conditions: 1) $\beta(0) =1$; 2) for all $t > 0$, $\beta(t)$ remains within the interval $[0, 1)$; and 3) all derivatives $\beta^{(i)}$ for $i = 0, 1, \dots, N$ are known, uniformly bounded, and piecewise continuous over $[0, \infty)$.

Based on the rate function, we construct the following scaling function
\begin{equation}\label{varphi}
   \varphi(t) = (\varphi_0 - \varphi_f )\beta(t)+\varphi_f
\end{equation}
where $ \varphi_f $ and $\varphi_0$ are some design parameters and satisfy $0 < \varphi_f < \varphi_0 \leq 1$. It can be verified that $\varphi(t)$ satisfies the following characteristics: 1) $\varphi(0) =\varphi_0 $ and $\varphi : (0,\infty) \rightarrow [\varphi_f , \varphi_0)$ for $t > 0$; and 2) all derivatives $\varphi^{(i)}$ for $i = 0,1,\dots,n$ are known, uniformly bounded, and piecewise continuous on $[0, \infty)$.

\begin{definition}
  A continuous function $\mathcal{H} : (-b_0,b_0) \rightarrow [-\infty , \infty)$ is called a intermediate function if 1) $\mathcal{H}$ is  continuously differentiable; 2) $\mathcal{H}(\varsigma)=-\mathcal{H}(-\varsigma)$, $\mathcal{H}(0)=0$, and $\lim_{\varsigma \rightarrow \pm b_0}\mathcal{H}(\varsigma)=\pm\infty$ ; and 3) $0<\upsilon\leq\dot{\mathcal{H}}$ and $\lim_{\varsigma \rightarrow \pm b_0}\dot{\mathcal{H}}(\varsigma)=\infty$, where $\upsilon$ is a constant.
\end{definition}

It is worth noting that there are many examples of such $\mathcal{H}$, e.g., $ \mathcal{H}(\varsigma)=\frac{\varsigma}{\sqrt{1-\varsigma^2}}$ and $\mathcal{H}(\varsigma)=\tan(\frac{\pi}{2}\varsigma)$ with $b_0=1$.

According to the definitions of $\varphi$ and $\mathcal{H}$, we call  $\mathcal{H}(\varphi)$ the performance function, and it is trivial to prove that $\mathcal{H}$ is a non-monotonic function as $\varphi$ is a non-monotonic function. To facilitate the presentation, this paper adopts the non-monotonic performance function as
\begin{equation}\label{pf}
 \mathcal{H}(\varphi)=\frac{l\varphi}{\sqrt{1-\varphi^2}}\ \textrm{with}\  \beta(t)=\exp(-\gamma t)\cos^2(t)
\end{equation}
where $l$ and $\gamma$ are some positive constants.
As noted in \cite{Bechlioulis2008tac,ZhangJ2020auto,LiZ2024smc}, the scaling function $\varphi(t)$ is commonly designed to decrease monotonically over time. However, \cite{Berger2021tac} and \cite{ZhaoK2023auto} have pointed out that employing a non-monotonic $\varphi(t)$, capable of temporarily relaxing the performance constraint, may offer practical benefits in specific scenarios.

\subsection{Error Transformation}
To achieve the performance constraint (\ref{ppb})) for $t\geq0$,  we define the error transformation function as
\begin{equation}\label{s1}
    {s}_j(t)=\frac{\zeta_j}{(\underline{\delta}_j+\zeta_j)(\bar{\delta}_j-\zeta_j)}, j=1,2,\ldots,n
\end{equation}
with $\zeta_j=\frac{\eta_j}{\varphi_j}$ and $\eta_j=\frac{e_j}{\sqrt{e_j^2+l_j^2}}$.

From (\ref{s1}), the variable $s_j$ exhibits the following characteristics: 1) for any $-\underline{\delta}_j < \zeta_j(0) < \bar{\delta}_j$, $s_j \rightarrow \pm\infty$ as $\zeta_j\rightarrow \bar{\delta}_j$ or $\zeta_j \rightarrow -\underline{\delta}_j$;
2) if $\zeta_j(0)$ satisfies $-\underline{\delta}_j < \zeta_j(0) < \bar{\delta}_j$ and $s_j(t)$ remains bounded for all $t \geq 0$, there exist tighter bounds $-\underline{\delta}_{1j}$ and $\bar{\delta}_{1j}$ within the original limits such that $-\underline{\delta}_j < -\underline{\delta}_{1j}\leq\zeta_j(0)\leq \bar{\delta}_{1j}< \bar{\delta}_j$.

The following lemma shall be used to explain how (\ref{ppb}) is ensured from (\ref{s1}).
\begin{lemma}\cite{ZhaoK2023auto}\label{l3}
If the control to be developed is able to guarantee the boundedness of $s_j(t)$ for $t \geq0$, then for any $\zeta_j(0)$ satisfying $-\underline{\delta}_j < \zeta_j(0) < \bar{\delta}_j$, (\ref{ppb}) is ensured.
\end{lemma}

\emph{{Proof:}} The detailed proof is provided in \cite{ZhaoK2023auto}, thus omitted.

The notable advantage of this error transformation is that the unified system framework for prescribed asymmetric and symmetric performance can be achieved by choosing the design parameters $\underline{\delta}_j$, $\bar{\delta}_j$ and $\varphi(0)$ (see \cite{ZhaoK2023auto} for details). Specifically, this error transformation framework can achieve global results similar to \cite{Berger2021tac,ZhangJ2021tac,ZhaoK2021tac,LiuG2024smc} (without considering error overshoot), as well as results similar to PPC \cite{Bechlioulis2008tac,HuangX2022tac} and asymmetric performance behaviors.

\subsection{Controller Design}\label{S33}
{To stabilize system (\ref{os}) with unified prescribed performance, here we develop a robust adaptive controller by applying the backstepping method in \cite{Krstic1995}. Let us begin the control design by defining the transformed tracking error and the virtual control errors as
\begin{align}
  {\bm{\varepsilon}}_1&=[{s}_1,\cdots,{s}_n]^T\label{z1}\\
  {\bm{\varepsilon}}_i&=\bm{x}_i-{\bm{a}}_{i-1},\ i=2,\ldots,N \label{zi}
\end{align}
where ${s}_j$ ($j=1,\ldots,n$) is defined in (\ref{s1}), ${\bm{a}}_{i-1}$ is a virtual controller to be designed at the $i-1$th step and the actual controller $\bm{u}$ is derived at the final step.}

{\emph{\textbf{Step 1:}} Differentiating (\ref{s1}) w.r.t. time, we have
\begin{equation}\label{ds1}
  \dot{s}_j=\mu_j\dot{\zeta}_j=\frac{\mu_jr_j}{\varphi_j}\dot{e}_j-\frac{\mu_j\dot{\varphi}_j\eta_j}{\varphi_j^2}=w_j(\dot{e}_j+v_j)
\end{equation}
where $\mu_j=\frac{\underline{\delta}_j\bar{\delta}_j+\zeta_j^2}{(\underline{\delta}_j+\zeta_j)^2(\bar{\delta}_j-\zeta_j)^2}$ is well-defined in the set $\Omega_{\zeta_j}=\{\zeta_j \in \mathbb{R} : -\underline{\delta}_j < \zeta_j < \bar{\delta}_j\}$, $r_j=\frac{l_j^2}{\sqrt{e_j^2+l_j^2}(e_j^2+l_j^2)}$,  $w_j=\frac{\mu_jr_j}{\varphi_j}$, and $v_j=-\frac{\dot{\varphi}_j\eta_j}{\varphi_jr_j}$. Further, (\ref{ds1}) can be written in compact form as
\begin{equation}\label{dS1}
  \dot{\bm{\varepsilon}}_1=\bm{W}(\dot{\bm{e}}+\bm{V})
\end{equation}
where $\bm{W}=\textrm{diag}\{w_j\}\in \mathbb{R}^{n \times n}$ and $\bm{V}=[v_1,\ldots,v_n]^T\in \mathbb{R}^{n}$.
Then, from the system (\ref{os}), we obtain
\begin{equation}\label{dz}
  \dot{\bm{\varepsilon}}_1=\bm{W}\left(\bm{f}_1+\bm{g}_1\bm{x}_{2}+{\bm{d}_1}-\dot{\bm{y}}_d+\bm{V}\right)
\end{equation}
where $\dot{\bm{\varepsilon}}_1=[\dot{s}_1,\cdots,\dot{s}_n]^T$. According to (\ref{s1}), $\bm{W}$ and $\bm{V}$ are available for control design, moreover, $\bm{W}$ is invertible and $\bm{W} > 0$ as long as $-\underline{\delta}_j < \zeta_j < \bar{\delta}_j$ over the interval $t\in[0,\tau_{\textrm{max}}]$. }

From (\ref{dz}) and (\ref{zi}), it is derived that
\begin{equation}\label{dz1}
  \dot{\bm{\varepsilon}}_1=\bm{W}\left(\bm{f}_1+\bm{g}_1{\bm{\varepsilon}}_{2}+\bm{g}_1{\bm{a}}_{1}+{\bm{d}_1}-\dot{\bm{y}}_d+\bm{V}\right)
\end{equation}
Next, consider the following Lyapunov candidate function:
\begin{equation}\label{v11}
  V_{11}=\frac{1}{2}{\bm{\varepsilon}}_1^T\bm{P}_1{\bm{\varepsilon}}_1
\end{equation}
 where $\bm{P}_1\in \mathbb{R}^{n\times n}$ is an unknown diagonal and positive definite matrix under \emph{Assumption} \ref{A2}.
Taking the time derivative of $V_{11}$ along (\ref{dz1}) yields
\begin{equation}
\begin{aligned}\label{dv111}
  \dot{V}_{11}=\ &{\bm{\varepsilon}}_1^T\bm{P}_1\bm{W}\left(\bm{f}_1+\bm{g}_1{\bm{\varepsilon}}_{2}+\bm{g}_1{\bm{a}}_{1}+{\bm{d}_1}-\dot{\bm{y}}_d+\bm{V}\right)\\
\ &+\frac{1}{2}{\bm{\varepsilon}}_1^T\dot{\bm{P}}_1{\bm{\varepsilon}}_1\\
  =\ &{\bm{\varepsilon}}_1^T\bm{W}\bm{P}_1\bm{g}_1{\bm{a}}_{1}+{\bm{\varepsilon}}_1^T\bm{W}\bm{P}_1\bm{g}_1{\bm{\varepsilon}}_{2}+\bm{h}_1
\end{aligned}
\end{equation}
with $\bm{h}_1={\bm{\varepsilon}}_1^T\bm{W}\bm{P}_1\left({\bm{f}}_1+{\bm{d}_1}-\dot{\bm{y}}_d+\bm{V}\right)+\frac{1}{2}{\bm{\varepsilon}}_1^T\dot{\bm{P}}_1{\bm{\varepsilon}}_1$
where the diagonal property of $\bm{W}$ and $\bm{P}_1$ is used.
Upon employing Young's inequality along with \emph{Assumption} \ref{A2} and \ref{A3}, one has
\begin{align*}
 {\bm{\varepsilon}}_1^T\bm{W}\bm{P}_1({\bm{f}}_1+{\bm{d}_1})\leq&\ \|\bm{W}\bm{\varepsilon}_1\|^2a_{11}^2a_{f1}^2\phi_{11}^2\phi_{f1}^2+\frac{1}{4}\\
  {\bm{\varepsilon}}_1^T\bm{W}\bm{P}_1(-\dot{\bm{y}}_d+\bm{V})\leq&\ \|\bm{W}\bm{\varepsilon}_1\|^2a_{11}^2\phi_{11}^2(\|\dot{\bm{y}}_d\|+\|\bm{V}\|)^2\\
  &+\frac{1}{4}.
\end{align*}
Since {$\frac{1}{2}{\bm{\varepsilon}}_1^T\dot{\bm{P}}_1{\bm{\varepsilon}}_1\leq \frac{1}{2}\|\bm{W}\bm{\varepsilon}_1\|^2a_{12}(\lambda_\textrm{min}(\bm{W}))^{-2}\phi_{12}$},
then the uncertain function $\bm{h}_1$ can be upper bounded by
\begin{equation}\label{ih1}
  \bm{h}_1\leq \|\bm{W}\bm{\varepsilon}_1\|^2\theta_1\Phi_1+\frac{1}{2}
\end{equation}
with $\theta_1=\max\{a_{11}^2a_{f1}^2,a_{11}^2,a_{12}\}$ being an unknown positive constant and {$\Phi_1=\phi_{11}^2\phi_{f1}^2+\phi_{11}^2(\|\dot{\bm{y}}_d\|+\|\bm{V}\|)^2+\frac{1}{2}(\lambda_\textrm{min}(\bm{W}))^{-2}\phi_{12}$ denoting a known and computable scalar function. Note that in $\bm{h}_1$, the potential destruction of the controllability condition by the performance incidental matrix $\bm{W}$ is neatly avoided by extending the ${\bm{\varepsilon}}_1$-based feedback to ($\bm{W}{\bm{\varepsilon}}_1$)-based feedback.}

Substituting (\ref{ih1}) into (\ref{dv111}), we have
\begin{equation}
\begin{aligned}\label{dv112}
  \dot{V}_{11}\leq&\ {\bm{\varepsilon}}_1^T\bm{W}\bm{P}_1\bm{g}_1({\bm{a}}_{1}
  +{\bm{\varepsilon}}_{2})+\|\bm{W}\bm{\varepsilon}_1\|^2\theta_1\Phi_1+\frac{1}{2}.
\end{aligned}
\end{equation}

At this stage, we construct the virtual controller $\bm{a}_1$ as:
\begin{align}
\bm{a}_1&=-\kappa_1\bm{W}{\bm{\varepsilon}}_1-{\hat{\theta}_1{\Phi_1}{\bm{W}{\bm{\varepsilon}}_1}}\label{u1}\\
\dot{\hat{\theta}}_1&={\sigma_1\|\bm{W}{\bm{\varepsilon}}_1\|^2}\Phi_1-\mu_1\hat{\theta}_1,\ \hat{\theta}_1(0)\geq0\label{a1}
\end{align}
where $\kappa_1>0$, $\mu_1>0$ and $\sigma_1>0$ are design parameters, $\hat{\theta}_1$ is the estimate of unknown constant $\theta_1$.
{We now introduce the complete Lyapunov function candidate as follows:
\begin{equation}
\begin{aligned}\label{v12}
V_1=V_{11}+\frac{1}{2\lambda_1\sigma_1}{\tilde{\theta}_1}^2
\end{aligned}
\end{equation}
where $\tilde{\theta}_1=\theta_1-\lambda_1\hat{\theta}_1$ is the parameter estimate error}.
Differentiating (\ref{v12}) and using (\ref{dv112})-(\ref{a1}) yields
\begin{equation}
\begin{aligned}\label{dv12}
  \dot{V}_{1}\leq& -{\bm{\varepsilon}}_1^T\bm{W}\bm{P}_1\bm{g}_1\left(\kappa_1\bm{W}{\bm{\varepsilon}}_1+{\hat{\theta}_1{\Phi_1}{\bm{W}{\bm{\varepsilon}}_1}}\right)+\frac{\mu_1}{\sigma_1}\tilde{\theta}_1\hat{\theta_1}\\
  &+(\theta_1-\tilde{\theta}_1){\|\bm{W}{\bm{\varepsilon}}_1\|^2\Phi_1}+{\bm{\varepsilon}}_1^T\bm{W}\bm{P}_1\bm{g}_1{\bm{\varepsilon}}_{2}+\frac{1}{2}.
\end{aligned}
\end{equation}
Note that
 \begin{equation}
\begin{aligned}\label{sgs11}
\bm{P}_1 \bm{g}_1=\, \frac{\bm{G}_1}{2}
+\frac{\left({\bm{P}_1 \bm{g}_1-\bm{g}_1^T\bm{P}_1}\right)}{2}
\end{aligned}
\end{equation}
where $\bm{G}_1$ is a symmetric matrix and ${\bm{P}_1 \bm{g}_1-\bm{g}_1^T\bm{P}_1}$ is a skew-symmetric matrix, then by \emph{Assumption} \ref{A2}, it holds that
 \begin{equation}
\begin{aligned}\label{sgs12}
&{\bm{\varepsilon}}_1^T\bm{W}\left({\bm{P}_1 \bm{g}_1-\bm{g}_1^T\bm{P}_1}\right)\bm{W}{\bm{\varepsilon}}_1=0,\\
&{\bm{\varepsilon}}_1^T\bm{W}\bm{G}_1\bm{W}{\bm{\varepsilon}}_1\geq\lambda_1\|\bm{W}{\bm{\varepsilon}}_1\|^2.
\end{aligned}
\end{equation}
Subsequently, inserting (\ref{sgs11}) and (\ref{sgs12}) into (\ref{dv12}) yields
\begin{equation}
\begin{aligned}\label{dv121}
  \dot{V}_{1}\leq& -\kappa_1\lambda_1\|\bm{W}{\bm{\varepsilon}}_1\|^2 - \lambda_1\hat{\theta}_1{\|\bm{W}{\bm{\varepsilon}}_1\|^2\Phi_1}+\frac{\mu_1}{\sigma_1}\tilde{\theta}_1\hat{\theta_1}\\
  &+(\theta_1-\tilde{\theta}_1){\|\bm{W}{\bm{\varepsilon}}_1\|^2\Phi_1}+\frac{1}{2}+{\bm{\varepsilon}}_1^T\bm{W}\bm{P}_1\bm{g}_1{\bm{\varepsilon}}_{2}\\
  \leq& -\kappa_1\lambda_1\|\bm{W}{\bm{\varepsilon}}_1\|^2-\frac{\mu_1}{2\sigma_1}\tilde{\theta}_1^2
  +\Delta_1+{\bm{\varepsilon}}_1^T\bm{W}\bm{P}_1\bm{g}_1{\bm{\varepsilon}}_{2}
\end{aligned}
\end{equation}
where the fact that $\frac{\mu_1}{\sigma_1}\tilde{\theta}_1\hat{\theta}_1\leq\frac{\mu_1}{2\sigma_1}{\theta}_1^2-\frac{\mu_1}{2\sigma_1}\tilde{\theta}_1^2$ is used, and $\Delta_1=\frac{\mu_1}{2\sigma_1}{\theta}_1^2+\frac{1}{2}$ is a positive constant.
The term ${\bm{\varepsilon}}_1^T\bm{W}\bm{P}_1\bm{g}_1{\bm{\varepsilon}}_{2}$ will be addressed in the next step.

\emph{\textbf{Step 2:}} Differentiating ${\bm{\varepsilon}}_2 = \bm{x}_2 - \bm{a}_1$ with respect to time, we obtain:
\begin{equation}\label{dz2}
  \dot{\bm{\varepsilon}}_2=\bm{f}_2+\bm{g}_2{\bm{\varepsilon}}_{3}+\bm{g}_2{\bm{a}}_{2}+{\bm{d}_2}-\dot{\bm{a}}_1
\end{equation}
where $\dot{\bm{a}}_1=\frac{\partial\bm{a}_1}{\partial\bm{x}_1}\left(\bm{f}_1
+\bm{g}_1\bm{x}_2+{\bm{d}_1}\right)+\bm{\omega}_1$ with $\bm{\omega}_1=\sum_{k=0}^{1}\big(\frac{\partial\bm{a}_1}{\partial\bm{y}_{d}^{(k)}}{\bm{y}}_{d}^{(k+1)}
+\frac{\partial\bm{a}_1}{\partial{\varphi}^{(k)}}{\varphi}^{(k+1)}\big)+\frac{\partial\bm{a}_1}{\partial{\hat{\theta}}_1}\dot{\hat{\theta}}_1$.

{Choose the Lyapunov candidate function as
$V_{21}=V_1+\frac{1}{2}{\bm{\varepsilon}}_2^T\bm{P}_2{\bm{\varepsilon}}_2$,
 where the matrix $\bm{P}_2 \in \mathbb{R}^{n \times n}$ is specified in \emph{Assumption} \ref{A2}.
Taking the time derivative of $V_{21}$ along (\ref{dz2}) yields
\begin{equation}
\begin{aligned}\label{dv21}
\dot{V}_{21}\leq&-\kappa_1\lambda_1\|\bm{W}{\bm{\varepsilon}}_1\|^2-\frac{\mu_1}{2\sigma_1}\tilde{\theta}_1^2
  +\Delta_1\\
&+{\bm{\varepsilon}}_2^T\bm{P}_2 \bm{g}_2{\bm{\varepsilon}}_{3}+{\bm{\varepsilon}}_2^T\bm{P}_2\bm{g}_2{\bm{a}}_{2}+ \bm{h}_2
\end{aligned}
\end{equation}
where $\bm{h}_2={\bm{\varepsilon}}_2^T \bm{P}_2\left(\bm{f}_2+{\bm{d}_2}-\frac{\partial\bm{a}_1}{\partial\bm{x}_1}\left(\bm{f}_1+{\bm{d}}_1+\bm{g}_1\bm{x}_2\right)-\bm{\omega}_1\right)+{\bm{\varepsilon}}_1^T\bm{W}\bm{P}_1\bm{g}_1{\bm{\varepsilon}}_{2}+\frac{1}{2}{\bm{\varepsilon}}_2^T \dot{\bm{P}_2}{\bm{\varepsilon}}_2$.
Similar to (\ref{ih1}), by using Young's inequality and \emph{Assumption} \ref{A3}, it is readily shown that the uncertain function $\bm{h}_2$ can be upper bounded by
\begin{equation}\label{ih2}
  \bm{h}_2\leq \|\bm{\varepsilon}_2\|^2\theta_2\Phi_2+\frac{5}{4}
\end{equation}
with $\theta_2=\max\{a_{21}^2a_{f2}^2,a_{21}^2a_{f1}^2,a_{21}^2\bar{g}_1^2,a_{21}^2,a_{11}^2\bar{g}_1^2,a_{22}\}$ being an unknown positive constant and $\Phi_2=\phi_{21}^2\phi_{f2}^2+\phi_{21}^2\phi_{f1}^2\|\frac{\partial\bm{a}_1}{\partial\bm{x}_1}\|^2
+\phi_{21}^2\|\frac{\partial\bm{a}_1}{\partial\bm{x}_1}\|^2
+\phi_{21}^2\|\bm{\omega}_1\|^2+\phi_{11}^2\|\bm{W}{\bm{\varepsilon}}_1\|^2
+\frac{1}{2}\phi_{22}$ denoting a known and computable scalar function.
Then substituting (\ref{ih2}) into (\ref{dv21}), it follows that
\begin{equation}
\begin{aligned}\label{dv212}
  \dot{V}_{21}\leq & -\kappa_1\lambda_1\|\bm{W}{\bm{\varepsilon}}_1\|^2-\frac{\mu_1}{2\sigma_1}\tilde{\theta}_1^2+{\bm{\varepsilon}}_2^T\bm{P}_2 \bm{g}_2{\bm{\varepsilon}}_{3}\\
&+{\bm{\varepsilon}}_2^T\bm{P}_2\bm{g}_2{\bm{a}}_{2}+ \|\bm{\varepsilon}_2\|^2\theta_2\Phi_2+\Delta_1+\frac{5}{4}.
\end{aligned}
\end{equation}

Now the virtual controller $\bm{a}_2$ is designed as:
\begin{align}
    \bm{a}_2&=-\kappa_2{\bm{\varepsilon}}_2-{\hat{\theta}_2{\Phi_2}{{\bm{\varepsilon}}_2}}\label{u2}\\
    \dot{\hat{\theta}}_2&={\sigma_2\|{\bm{\varepsilon}}_2\|^2}\Phi_2-\mu_2\hat{\theta}_2,\ \hat{\theta}_2(0)\geq0\label{a2}
\end{align}
where $\kappa_2>0$, $\mu_2>0$ and $\sigma_2>0$ are design parameters, $\hat{\theta}_2$ is the estimate of $\theta_2$.

{Constructing the complete Lyapunov function candidate as
$V_2=V_{21}+\frac{1}{2\lambda_2\sigma_2}{\tilde{\theta}_2}^2$, where $\tilde{\theta}_2=\theta_2-\lambda_2\hat{\theta}_2$ is the estimate error.}
Then, differentiating $V_2$ and using (\ref{dv212})-(\ref{a2}) yields
\begin{equation}
\begin{aligned}\label{dv22}
\dot{V}_{2}\leq & -\kappa_1\lambda_1\|\bm{W}{\bm{\varepsilon}}_1\|^2-\frac{\mu_1}{2\sigma_1}\tilde{\theta}_1^2 +\lambda_2\hat{\theta}_2{{\|{\bm{\varepsilon}}_2\|^2\Phi_2}}\\ &+\frac{\mu_2}{\sigma_2}\tilde{\theta}_2\hat{\theta}_2-{\bm{\varepsilon}}_2^T\bm{P}_2\bm{g}_2\left(\kappa_2{\bm{\varepsilon}}_2+{\hat{\theta}_2{\Phi_2}{{\bm{\varepsilon}}_2}}\right)
\\
&+{\bm{\varepsilon}}_2^T\bm{P}_2\bm{g}_2{\bm{\varepsilon}}_{3}+\Delta_1+\frac{5}{4}.
\end{aligned}
\end{equation}
Similar to the analysis in (\ref{sgs11}) and (\ref{sgs12}), we have that
 \begin{equation}\label{sgs22}
{\bm{\varepsilon}}_2^T\bm{P}_2\bm{g}_2(\kappa_2{\bm{\varepsilon}}_2
+{\hat{\theta}_2{\Phi_2}{{\bm{\varepsilon}}_2}})\geq(\kappa_2\lambda_2+\lambda_2{\hat{\theta}_2{\Phi_2}})\|{\bm{\varepsilon}}_2\|^2
\end{equation}
which further leads to
\begin{equation}
\begin{aligned}\label{dv221}
  \dot{V}_{2}\leq & -\kappa_1\lambda_1\|\bm{W}{\bm{\varepsilon}}_1\|^2 -\kappa_2\lambda_2\|{\bm{\varepsilon}}_2\|^2-\sum_{k=1}^{2}\frac{\mu_k}{2\sigma_k}\tilde{\theta}_k^2\\ &+\Delta_2+{\bm{\varepsilon}}_2^T\bm{P}_2\bm{g}_2{\bm{\varepsilon}}_{3}
\end{aligned}
\end{equation}
where the fact that $\frac{\mu_2}{\sigma_2}\tilde{\theta}_2\hat{\theta}_2\leq\frac{\mu_2}{2\sigma_2}{\theta}_2^2-\frac{\mu_2}{2\sigma_2}\tilde{\theta}_2^2$ is used, and $\Delta_2=\Delta_1+\frac{\mu_2}{2\sigma_2}{\theta}_2^2+\frac{5}{4}$ is a positive constant. The term ${\bm{\varepsilon}}_2^T\bm{P}_2\bm{g}_2{\bm{\varepsilon}}_{3}$ will be handled in next step.

\emph{\textbf{Step i} ($i=3,\ldots,N-1$):} Based on the first two steps and considering the Lyapunov function candidate as
\begin{equation}\label{vi}
  V_{i}=V_{i-1}+\frac{1}{2}{\bm{\varepsilon}}_i^T\bm{P}_i{\bm{\varepsilon}}_i+\frac{1}{2\lambda_i\sigma_i}{\tilde{\theta}_i}^2,\ i=3,\ldots,N-1
\end{equation}
where $\bm{P}_i\in \mathbb{R}^{n\times n}$ is defined in \emph{Assumption} \ref{A2}, and ${\tilde{\theta}_i}=\theta_i-\lambda_i\hat{\theta}_i$. {It is worth noting that the auxiliary matrix $\bm{P}_i$ is $(\bm{X}_{i-1},t)$-dependent rather than $(\bm{X}_{i},t)$-dependent, because $\bm{P}_i$ being $(\bm{X}_{i},t)$-dependent would cause $\partial \bm{P}_{i}/\partial t$ to be $(\bm{X}_{i+1},t)$-dependent, and thus the virtual controller $\bm{\alpha}_{i}$ will be $(\bm{X}_{i+1},t)$-dependent, which will lead to the algebraic loop problem in the recursive design procedure.}

The virtual controllers $\bm{a}_i$ ($i=3,\ldots,N-1$) and adaptive laws $\dot{\hat{\theta}}_i$ ($i=3,\ldots,N-1$) can be recursively obtained by following the standard backstepping procedure, as summarized in Table \ref{table 1}. Note that $\kappa_i>0$, $\mu_i>0$ and $\sigma_i>0$ are design parameters, $\hat{\theta}_i$ is the estimate of $\theta_i$ with $\theta_i=\max\{a_{i1}^2a_{fi}^2,a_{i1}^2a_{f(i-1)}^2,a_{i1}^2\bar{g}_{i-1}^2,a_{i1}^2,a_{(i-1)1}^2\bar{g}_{i-1}^2,a_{i2}\}$.

\begin{table}[!htbp]
	\caption{Adaptive backstepping virtual controller.}
	\rule[1pt]{8.85cm}{0.1em}
	\textbf{Virtual Control Schemes: }($i=3,\ldots,N-1$)\\
	\rule[1pt]{8.85cm}{0.05em}
\begin{flalign}
&\ \bm{a}_i=-\kappa_i{\bm{\varepsilon}}_i-{\hat{\theta}_i{\Phi_i}{{\bm{\varepsilon}}_i}}&\label{ui}\\
&\ \dot{\hat{\theta}}_i={\sigma_i\|{\bm{\varepsilon}}_i\|^2}\Phi_i-\mu_i\hat{\theta}_i,\ \hat{\theta}_i(0)\geq0&\label{ai}
\end{flalign}
with $\Phi_i=\phi_{i1}^2\phi_{fi}^2+\phi_{i1}^2\phi_{f(i-1)}^2\|\frac{\partial\bm{a}_{i-1}}{\partial\bm{x}_{i-1}}\|^2
+\phi_{i1}^2\|\frac{\partial\bm{a}_{i-1}}{\partial\bm{x}_{i-1}}\|^2 +\phi_{i1}^2\|\bm{\omega}_{i-1}\|^2\\+\phi_{(i-1)1}^2\|{\bm{\varepsilon}}_{i-1}\|^2+\frac{1}{2}\phi_{i2}$,
\\ $\bm{\omega}_{i-1}=\sum_{k=0}^{i-1}\big(\frac{\partial\bm{a}_{i-1}}{\partial\bm{y}_{d}^{(k)}}{\bm{y}}_{d}^{(k+1)}
+\frac{\partial\bm{a}_{i-1}}{\partial{\varphi}^{(k)}}{\varphi}^{(k+1)}\big)+\sum_{k=1}^{i-1}\frac{\partial\bm{a}_{i-1}}{\partial{\hat{\theta}}_{k}}\dot{\hat{\theta}}_k$.\\
	\rule[1pt]{8.85cm}{0.05em}
\label{table 1}
\end{table}

Similar to the analysis in \emph{\textbf{Step 2}}, it is directly deduced from Table \ref{table 1} that
\begin{equation}
\begin{aligned}\label{dvi1}
  \dot{V}_{i}\leq & -\kappa_1\lambda_1\|\bm{W}{\bm{\varepsilon}}_1\|^2 -\sum_{k=2}^{i}\kappa_{k}\lambda_k\|{\bm{\varepsilon}}_k\|^2\\ &-\sum_{k=1}^{i}\frac{\mu_k}{2\sigma_k}\tilde{\theta}_k^2+\Delta_i+{\bm{\varepsilon}}_i^T\bm{P}_i\bm{g}_i{\bm{\varepsilon}}_{i+1}
\end{aligned}
\end{equation}
where $\Delta_i=\Delta_{i-1}+\frac{\mu_i}{2\sigma_i}{\theta}_i^2+\frac{5}{4}$ is a positive constant, and ${\bm{\varepsilon}}_i^T\bm{P}_i\bm{g}_i{\bm{\varepsilon}}_{i
+1}$ will be handled in final step.

\emph{\textbf{Step N} :} From (\ref{os}) and (\ref{zi}), the time derivative of $\bm{\varepsilon}_N$ can be obtained as
\begin{equation}\label{dzn}
  \dot{\bm{\varepsilon}}_N=\bm{f}_N+\bm{g}_N\bm{\rho}\bm{u}+\bm{g}_N\bm{v}+{\bm{d}_N}-\dot{\bm{a}}_{N-1}
\end{equation}
where
$\dot{\bm{a}}_{N-1}=\sum_{k=1}^{N-1}\frac{\partial\bm{a}_{N-1}}{\partial\bm{x}_{k}}\left(\bm{f}_{k}
+{\bm{d}_k}+\bm{g}_{k}\bm{x}_{k+1}\right)+\bm{\omega}_{N-1}$ with
$\bm{\omega}_{N-1}=\sum_{k=0}^{N-1}\big(\frac{\partial\bm{a}_{N-1}}{\partial\bm{y}_{d}^{(k)}}{\bm{y}}_{d}^{(k+1)}
+\frac{\partial\bm{a}_{N-1}}{\partial{\varphi}^{(k)}}{\varphi}^{(k+1)}\big)+\sum_{k=1}^{N-1}\frac{\partial\bm{a}_{N-2}}{\partial{\hat{\theta}}_{k}}\dot{\hat{\theta}}_k$.

Then considering the Lyapunov function candidate as
\begin{equation}\label{vn1}
  V_{N1}=V_{N-1}+\frac{1}{2}{\bm{\varepsilon}}_N^T\bm{P}_{N}{\bm{\varepsilon}}_N
\end{equation}
where $\bm{P}_{N}\in \mathbb{R}^{n\times n}$ is specified in \emph{Assumption} \ref{A2}.

Differentiating $V_{N1}$ with respect to time along (\ref{dzn}) and applying \emph{Assumption} \ref{A2}, we obtain:
\begin{equation}
\begin{aligned}\label{dvn1}
   \dot{V}_{N1}\leq& -\kappa_1\lambda_1\|\bm{W}{\bm{\varepsilon}}_1\|^2 -\sum_{k=2}^{N-1}\kappa_{k}\lambda_k\|{\bm{\varepsilon}}_k\|^2-\sum_{k=1}^{N-1}\frac{\mu_k}{2\sigma_k}\tilde{\theta}_k^2\\ &+\Delta_{N-1}+{\bm{\varepsilon}}_N^T\bm{\mathcal{G}}_N{\bm{u}}+ \bm{h}_N
\end{aligned}
\end{equation}
where $\bm{\mathcal{G}}_N=\bm{P}_N\bm{g}_N\bm{\rho}$ and $\bm{h}_N={\bm{\varepsilon}}_N^T\bm{P}_{N}\big(\bm{f}_N+{\bm{d}_N}-\sum_{k=1}^{N-1}
 \frac{\partial\bm{a}_{N-1}}{\partial\bm{x}_{k}}\left(\bm{f}_{k}
+{\bm{d}_k}+\bm{g}_{k}\bm{x}_{k+1}\right)-\bm{\omega}_{N-1}
+\bm{g}_{N}\bm{\upsilon}\big)+{\bm{\varepsilon}}_{N-1}^T\bm{P}_{N-1}\bm{g}_{N-1}{\bm{\varepsilon}}_{N}+\frac{1}{2}{\bm{\varepsilon}}_N^T\dot{\bm{P}}_{N}{\bm{\varepsilon}}_N$.
Similar to \emph{\textbf{Step i}} ($i=1, \ldots, N-1$), the uncertain function $\bm{h}_N$ can be upper bounded by
\begin{equation}\label{ihn}
\bm{h}_N\leq \|\bm{\varepsilon}_N\|^2\theta_N\Phi_N+\frac{3}{2}
\end{equation}
with $\theta_N=\max\{{a}_{N1}^2a_{fN}^2,{a}_{N1}^2a_{f(N-1)}^2,{a}_{N1}^2\bar{g}_{N-1}^2,{a}_{N1}^2\bar{g}_N^2\bar{\upsilon}^2,\\{a}_{N1}^2,a_{(N-1)1}^2\bar{g}_{N-1}^2,{a}_{N2}\}$ being an unknown positive constant and $\Phi_N={\phi}_{N1}^2\phi_{fN}^2+{\phi}_{N1}^2\phi_{f(N-1)}^2\|\frac{\partial\bm{a}_{N-1}}{\partial\bm{x}_{N-1}}\|^2
+{\phi}_{N1}^2\|\frac{\partial\bm{a}_{N-1}}{\partial\bm{x}_{N-1}}\|^2+{\phi}_{N1}^2\|\bm{\omega}_{N-1}\|^2+{\phi}_{N1}^2+\phi_{(N-1)1}^2\|{\bm{\varepsilon}}_{N-1}\|^2
+\frac{1}{2}{\phi}_{N2}$ denoting a known and computable scalar function.

By inserting (\ref{ihn}) into (\ref{dvn1}), we can get
\begin{equation}
\begin{aligned}\label{dvn12}
  \dot{V}_{N1}\leq & -\kappa_1\lambda_1\|\bm{W}{\bm{\varepsilon}}_1\|^2 -\sum_{k=2}^{N-1}\kappa_{k}\lambda_k\|{\bm{\varepsilon}}_k\|^2-\sum_{k=1}^{N-1}\frac{\mu_k}{2\sigma_k}\tilde{\theta}_k^2\\ &+\Delta_{N-1}+{\bm{\varepsilon}}_N^T\bm{\mathcal{G}}_N{\bm{u}}+ \|\bm{\varepsilon}_N\|^2\theta_N\Phi_N+\frac{3}{2}.
\end{aligned}
\end{equation}

The actual control law $u$ is designed as:
\begin{align}
\bm{u}&=-\frac{\bm{A}^T}{\|\bm{A}\|}\left(\kappa_N{\bm{\varepsilon}}_N+{\hat{\theta}_N{\Phi_N}{{\bm{\varepsilon}}_N}}\right)\label{un}\\
\dot{\hat{\theta}}_N&={\sigma_N\|{\bm{\varepsilon}}_N\|^2}\Phi_N-\mu_N\hat{\theta}_N,\ \hat{\theta}_N(0)\geq0\label{an}
\end{align}
where $\kappa_N>0$, $\mu_N>0$ and $\sigma_N>0$  are  design parameters, $\hat{\theta}_N$ is the estimate of $\theta_N$.

\begin{remark}
{The controllability condition constructed in this paper has several merits in solving the unknown yet time-varying control gain matrix $\bm{g}_i$: 1) in contrast to \cite{Bechlioulis2008tac,YangQ2015auto}, no approximation methods are used to estimate $\bm{g}_i$, thus avoiding the singularity problem \cite{ZhangY2019auto} that may arise when calculating the inverse of the estimated matrix; 2) it does not impose assumptions on $\bm{g}_i$ directly or decompose the matrix $\bm{L}$ offline to satisfy the controllability requirement, but rather, it satisfies structural assumption through the introduction of auxiliary matrix $\bm{P}_i$, which thus does not require any information about $\bm{g}_i$, and hence, can be applied to a wider variety of practical systems; and 3) avoiding using $\bm{\rho}$ to construct the Liapunov function when dealing with multiplicative actuator faults as used in \cite{ZhangJ2020auto,ZhaoK2023auto}, thereby allowing for handling the more generalized intermittent faults where $\bm{\rho}$ is not satisfied to be continuous and derivable w.r.t. time.}
\end{remark}

\begin{remark}\label{R6}
Although the auxiliary matrix $\bm{P}_i$ appears in the Lyapunov function $V_i$, it is not directly used in the control laws (\ref{u1}), (\ref{u2}), (\ref{ui}) and (\ref{un}). Instead, an adaptive mechanism based on “core function”-driven estimators is embedded into the backstepping design to implicitly handle the uncertainty in $\bm{P}_i$ and system nonlinearities. As a result, the proposed controllability conditions are satisfied online without requiring any feasibility assumptions on $\bm{P}_i$ or offline computation of auxiliary matrices such as $\bm{L}$, making the method more general than those in \cite{XuH2003tac,SongY2016tac,Lee2016scl,Katsoukis2021tac,HuangX2022tac,ZhouB2023tac}.
\end{remark}

\section{Stability Analysis}
 Now we give the stability analysis of the control laws designed in Section \ref{S33} for system (\ref{os}).

\begin{theorem}\label{the1}
Consider the uncertain non-square MIMO nonlinear strict-feedback system (\ref{os}) subject to actuator faults. Under \emph{Assumptions} \ref{A1}–\ref{A3}, if the control law (\ref{un}) is implemented together with the virtual controllers given in (\ref{u1}), (\ref{u2}), and (\ref{ui}), then all signals in the closed-loop system remain bounded despite the presence of actuator faults, and the tracking error $\bm{e}_j$ is guaranteed to stay within the prescribed performance bounds ($\mathcal{H}(-\varphi_j(t)),\mathcal{H}(\varphi_j(t))$).
\end{theorem}

\emph{\textbf{Proof:}}
First, we define $\tilde{\theta}_N=\theta_N-\lambda_N\hat{\theta}_N$, and then, blend such error into the complete Lyapunov function candidate such that
\begin{equation}\label{vn}
  V_{N}=V_{N1}+\frac{1}{2\lambda_N\sigma_N}{\tilde{\theta}_N}^2
\end{equation}
Differentiating (\ref{vn}) and using (\ref{un}) yields
\begin{equation}
\begin{aligned}\label{dvn21}
  \dot{V}_{N}\leq & -\kappa_1\lambda_1\|\bm{W}{\bm{\varepsilon}}_1\|^2 -\sum_{k=2}^{N-1}\kappa_{k}\lambda_k\|{\bm{\varepsilon}}_k\|^2-\sum_{k=1}^{N-1}\frac{\mu_k}{2\sigma_k}\tilde{\theta}_k^2\\ &-\frac{{\bm{\varepsilon}}_N^T\bm{\mathcal{G}}_N\bm{A}^T\left(\kappa_N{\bm{\varepsilon}}_N+{\hat{\theta}_N{\Phi_N}{{\bm{\varepsilon}}_N}}\right)}{\|\bm{A}\|}\\
  &+ \|\bm{\varepsilon}_N\|^2\theta_N\Phi_N-\frac{1}{\sigma_N}{\tilde{\theta}_N}{\dot{\hat{\theta}}_N}+\Delta_{N-1}+\frac{3}{2}.
\end{aligned}
\end{equation}
Note that
 \begin{equation}
\begin{aligned}\label{sgsN1}
\bm{\mathcal{G}}_N\bm{A}^T=\, \frac{\bm{G}_{N}}{2}+\frac{\left(\bm{\mathcal{G}}_N\bm{A}^T-\bm{A}\bm{\mathcal{G}}_N^T\right)}{2}
\end{aligned}
\end{equation}
where $\bm{G}_{N}$ is a symmetric matrix and $\bm{\mathcal{G}}_N\bm{A}^T-\bm{A}\bm{\mathcal{G}}_N^T$ is a skew-symmetric matrix, then by \emph{Assumption} \ref{A2}, it holds that
 \begin{equation}
\begin{aligned}\label{sgsn2}
&{\bm{\varepsilon}}_N^T\left(\bm{\mathcal{G}}_N\bm{A}^T-\bm{A}\bm{\mathcal{G}}_N^T\right){\bm{\varepsilon}}_N=0,\\
&{\bm{\varepsilon}}_N^T\bm{G}_{N}{\bm{\varepsilon}}_N\geq\lambda_N\|\bm{A}\|\|{\bm{\varepsilon}}_N\|^2.
\end{aligned}
\end{equation}
Then, substituting (\ref{sgsn2}) and adaptation law (\ref{an}) into (\ref{dvn21}) yields
\begin{equation}
\begin{aligned}\label{dvn22}
 \dot{V}_{N}\leq & -\kappa_1\lambda_1\|\bm{W}{\bm{\varepsilon}}_1\|^2 -\sum_{k=2}^{N}\kappa_{k}\lambda_k\|{\bm{\varepsilon}}_k\|^2-\sum_{k=1}^{N}\frac{\mu_k}{2\sigma_k}\tilde{\theta}_k^2\\
&+\Delta_{N-1}+\frac{\mu_N}{2\sigma_N}{\theta}_N^2+\frac{3}{2}\\
\leq &-\Upsilon V_N+C
\end{aligned}
\end{equation}
where $\Upsilon=\min\{\frac{2\lambda_{\textrm{min}}(\bm{W}^T\bm{W})\kappa_1\lambda_1}{\lambda_{\textrm{max}}(\bm{P}_{1})},\frac{2\kappa_k\lambda_k}{\lambda_{\textrm{max}}(\bm{P}_{N})},\frac{\mu_i}{\lambda_i}\}>0$ for $k=2,\ldots,N$, $i=1,\ldots,N$, and $C=\Delta_{N-1}+\frac{\mu_N}{2\sigma_N}{\theta}_N^2+\frac{3}{2}>0$.

We first establish the boundedness of all signals in the closed-loop system. From (\ref{dvn22}), we have $V_N(t)\leq V_N(0)+\frac{C}{\Upsilon}=\Xi$,
which implies that $V_N(t)$ remains bounded for all $t \in [0, \tau_{\textrm{max}})$. By the definition of $V_N(t)$, it follows that both ${\bm{\varepsilon}}_i$ and $\tilde{\theta}_i$ are bounded and satisfy ${\bm{\varepsilon}}_1\leq \sqrt{\frac{2\lambda_{\textrm{min}}(\bm{W}^T\bm{W})\Xi}{\lambda_{\textrm{max}}(\bm{P}_{1})}}$, ${\bm{\varepsilon}}_k\leq \sqrt{\frac{2\Xi}{\lambda_{\textrm{max}}(\bm{P}_{k})}}$ ($k=2,\ldots,N$), and $\tilde{\theta}_i\leq\sqrt{2\sigma_i\Xi}$ for $t \in[0,\tau_{\textrm{max}})$ with ${\lambda_{\textrm{max}}(\bm{P}_{k})}>0$, which further indicates that the parameter estimate $\hat{\theta}_i$ is bounded over the interval $t \in[0,\tau_{\textrm{max}})$. Since ${\bm{\varepsilon}}_1$ is bounded, and using the properties of $s_j$, it is shown that $-1\leq-\underline{\delta}_j < -\underline{\delta}_{1j}\leq\zeta_j(t)\leq \bar{\delta}_{1j}< \bar{\delta}_j\leq1$ for $t \in[0,\tau_{\textrm{max}})$. As $\zeta_j=\frac{\eta_j}{\varphi_j}$ and $0 < \varphi_{jf} < \varphi_j(t)\leq\varphi_{j0} \leq 1$, it is proved that there exist some constants $\underline{\eta}_j$ and $\bar{\eta}_j$ so that $-1 <\underline{\eta}_j \leq\eta_j \leq \bar{\eta}_j < 1$, which further implies from $e_j = \frac{l_j\eta_j}{\sqrt{1-\eta_j^2}}$ that the tracking error $e_j$ is bounded for $t \in[0,\tau_{\textrm{max}})$. As $e_j=x_{1j}-y_{dj}$, one has $x_{1j}$ is bounded since $y_{dj}$ is bounded for $t \in[0,\tau_{\textrm{max}})$. Hence,  it is trivial to prove that $0 < \underline{\mu}_j \leq \mu_j \leq\bar{\mu}_j$ and $0 < \underline{r}_j \leq r_j \leq\bar{r}_j$, $0 < \underline{w}_j \leq w_j \leq\bar{w}_j$ for $t \in[0,\tau_{\textrm{max}})$, with $\underline{\mu}_j$, $\bar{\mu}_j$, $\underline{r}_j$, $\bar{r}_j$,$\underline{w}_j$, $\bar{w}_j$ being some positive constants, which indicates that $\bm{W}$ and $\bm{V}$ are bounded. Since $x_{1}$ is bounded, it is shown from \emph{Assumption} \ref{A1} and \ref{A3} that $\Phi_1$ is bounded, and thus $\bm{a}_1$ and $\dot{\hat{\theta}}_1$ are bounded by (\ref{u1}) and (\ref{a1}). Following this line of reasoning, $\bm{a}_i$ ($i=2,\ldots,N-1$), $\bm{u}$, $\dot{\hat{\theta}}_k$ ($k=2,\ldots,N$) and $\bm{x}_k$ are bounded on $t \in[0,\tau_{\textrm{max}})$. Importantly, the boundedness of these signals depends solely on the design parameters and is independent of $\tau_\textrm{max}$, implying that $\tau_{\textrm{max}}$ can be extended to infinity. Therefore, all signals in the closed-loop system remain bounded for all $t \in [0, \infty)$.

We now demonstrate that the tracking error remains within the prescribed bounds for all $t \in [0, \infty)$. Since $s_j$ is bounded over this interval, it follows from Lemma \ref{l3} that the tracking error satisfies $\mathcal{H}(-\underline{\delta}_j\varphi_j(t)) < e_j(t) < \mathcal{H}(\bar{\delta}_j\varphi_j(t))$ for all $j=1,2,\ldots,m$. This completes the proof.
\QEDB

\begin{remark}\label{R7}
By introducing the unified prescribed performance function $\mathcal{H}$, the originally constrained error dynamics $\dot{\bm{e}}$ are converted into an unconstrained transformed system as shown in (\ref{dz}). The proposed virtual controllers (\ref{u1}), (\ref{u2}), (\ref{ui}) together with the actual controller (\ref{un}) then stabilize this transformed system, which indirectly enforces the predefined performance constraints on the tracking error. Notably, suitable selection of design parameters enables global or semi-global asymmetric performance to be realized without altering the control framework. Moreover, unexpected actuator faults are compensated automatically without the need for additional fault detection or diagnosis mechanisms.
\end{remark}

\begin{remark}
The controllability condition and controller constructed in this paper can be flexibly applied to a class of square and non-square systems. Specifically, square and non-square systems can be handled by simply choosing $\bm{A}=\bm{I}_n$ and $\bm{A}=[\bm{I}_n, \bm{\Lambda}_1,\ldots,\bm{\Lambda}_{(m-n)}]$, respectively. Moreover, since the auxiliary matrix $\bm{P}_i$ is adaptively compensated and does not use in the controller, even if $\bm{g}_i$ satisfies the original strong controllability condition in \cite{SongY2016tac}, it is sufficient to compensate $\bm{P}_i$ as a identity matrix $\bm{I}_n$ without any additional operation. 
Notably, the relaxed controllability condition proposed in this work is practically meaningful, as demonstrated in spacecraft applications \cite{ShaoX2023auto}. Although it does not alter the structure of existing controllers (e.g., \cite{SongY2016tac,LiX2023tac1}), it is clear from the preceding analysis that introducing the unknown auxiliary matrix $\bm{P}_i$ can significantly affect both controller design and stability analysis. In this work, such an unknown component is cleverly integrated with system uncertainties, allowing the controller to retain a unified and general structure, while extending applicability to more general and uncertain systems.
\end{remark}

\section{ Illustrative Example}\label{s5}
\subsection{Application to quadrotors (square case)}\label{s51}
%
In this section, we consider a quadrotor \cite{ZhouB2023tac} whose 3-DOF rotational dynamics are described by
\begin{equation}\label{afs2}
\begin{aligned}
\dot{\bm{\Theta}}=\,&\bm{R}\bm{Q}\\
\bm{M}\dot{\bm{Q}}=\,&-\bm{Q}\times\bm{M}\bm{Q}+\bm{\tau}+\bm{\Delta}(t)
\end{aligned}\end{equation}
where $\bm{\Theta} = [\phi, \theta, \psi]^T \in \mathbb{R}^3$ represents the Euler angles, $\bm{Q} = [p, q, r]^T \in \mathbb{R}^3$ denotes the angular velocity, and $\bm{\Delta}(t) \in \mathbb{R}^3$ is the disturbance vector. The control torque vector is given by $\bm{\tau} = [\tau_1, \tau_2, \tau_3]^T \in \mathbb{R}^3$, the inertia matrix is $\bm{M} = \operatorname{diag}\{M_{xx}, M_{yy}, M_{zz}\} \in \mathbb{R}^{3 \times 3}$, and $\bm{R} \in \mathbb{R}^{3 \times 3}$ denotes the transformation matrix. Further details on this model can be found in \cite{TianB2019tie}.
Defining $\bm{x}_1 = \bm{\Theta}$, $\bm{x}_2 = \dot{\bm{\Theta}}$, and $\bm{X}_2 = [\bm{x}_1^T, \bm{x}_2^T]^T$, the system (\ref{afs2}) can be equivalently rewritten as
\begin{equation}\label{afs3}
\begin{aligned}
\dot{\bm{x}}_{1}=\,&\bm{x}_2\\
\dot{\bm{x}}_{2}=\,&\bm{f}_2(\bm{X}_2)+\bm{g}_2(\bm{x}_1)\bm{u}_a+\bm{d}_2(\bm{x}_1,t)
\end{aligned}\end{equation}
with $\bm{f}_1(\cdot)=\bm{0}_3$, $\bm{f}_2(\cdot)=-\bm{R}\bm{M}^{-1}\bm{R}^{-1}\bm{x}_2\times\bm{M}\bm{R}^{-1}\bm{x}_2+\dot{\bm{R}}\bm{R}^{-1}\bm{x}_2$, $\bm{g}_1(\cdot)=\bm{I}_3$, $\bm{u}_a=\bm{\tau}$, $\bm{d}_1(\cdot)=\bm{0}_3$, $\bm{d}_2(\cdot)=\bm{R}\bm{M}^{-1}\bm{\Delta}(t)$, and $\bm{g}_2(\cdot)=\bm{R}\bm{M}^{-1}$ calculated as
\begin{equation}\label{R}
  \bm{g}_2(\bm{x}_1)=\begin{bmatrix}
    \frac{1}{M_{xx}} & \frac{\sin(\phi)\tan(\theta)}{M_{yy}} & \frac{\cos(\phi)\tan(\theta)}{M_{zz}} \\
    0 & \frac{\cos(\phi)}{M_{yy}} & \frac{ -\sin(\phi)}{M_{zz}} \\
    0 & \frac{\sin(\phi)\sec(\theta)}{M_{yy}}&\frac{\cos(\phi)\sec(\theta)}{M_{zz}}
  \end{bmatrix}.
\end{equation}
The fault profiles are outlined as follows:
  \begin{equation}
  \begin{aligned}
      &\bm{\rho}(t)=\begin{cases}\textmd{diag}\big\{1-0.1\sin(t),1-0.2\tanh(t),\\
      \qquad\ 0.9+0.1\cos(t)\big\},\ \ \ \ \ \, t\in(0,3] \\
      \textmd{diag}\big\{0.8+0.05\sin(t),0.8+0.02\tanh(0.5t),\\
      \qquad \  0.8+0.2\cos(t)\big\},\ \ t\in(3,\infty)\end{cases}\\
     & \bm{\upsilon}(t)= [0.02\tanh(2t);0.02\cos(t);0.02\sin(3t)].
  \end{aligned}
  \end{equation}
The quadrotor’s physical parameters are given by $M_{xx}=0.021\textrm{kg}\cdot\textrm{m}^2$, $M_{yy}=0.021\textrm{kg}\cdot\textrm{m}^2$, and $M_{zz}=0.039\textrm{kg}\cdot\textrm{m}^2$.
The desired reference trajectory is $\bm{y}_d=[\cos(t),\sin(2t),0.1\tanh(t)]^T$, and the disturbance vector is defined as $\bm{\Delta}(t)=[0.02\sin(t),0.02\cos(t),\tanh(t)]^T$.
For this square system, the actuator allocation matrix is the identity, $\bm{A} = \bm{I}_n$. Although the matrix $\bm{\rho}\bm{g}_2 + \bm{g}_2^T \bm{\rho}$ is not uniformly positive/negative definite for certain states $\bm{x}_1$, the system still meets Assumption \ref{A2} by selecting $\bm{P}_2=\textrm{diag}\{M_{xx}+\cos(\phi)^2,M_{yy}+\sec(\theta),M_{zz}+\frac{M_{yy}\cos(\theta)}{M_{zz}\cos(\phi)}\}$.
 \begin{figure}[htp]
      \centering
       {\includegraphics[width=6cm]{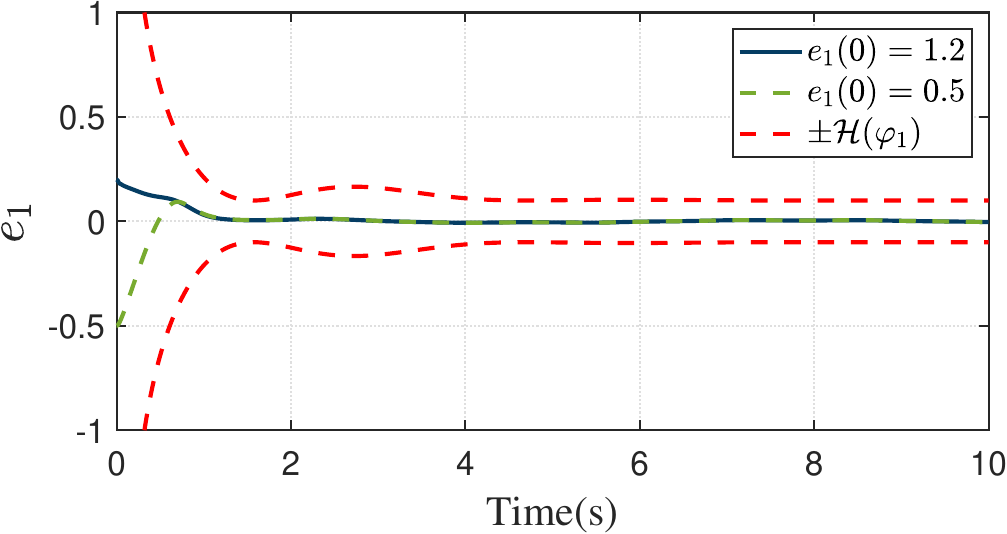}}\\
       {\includegraphics[width=6cm]{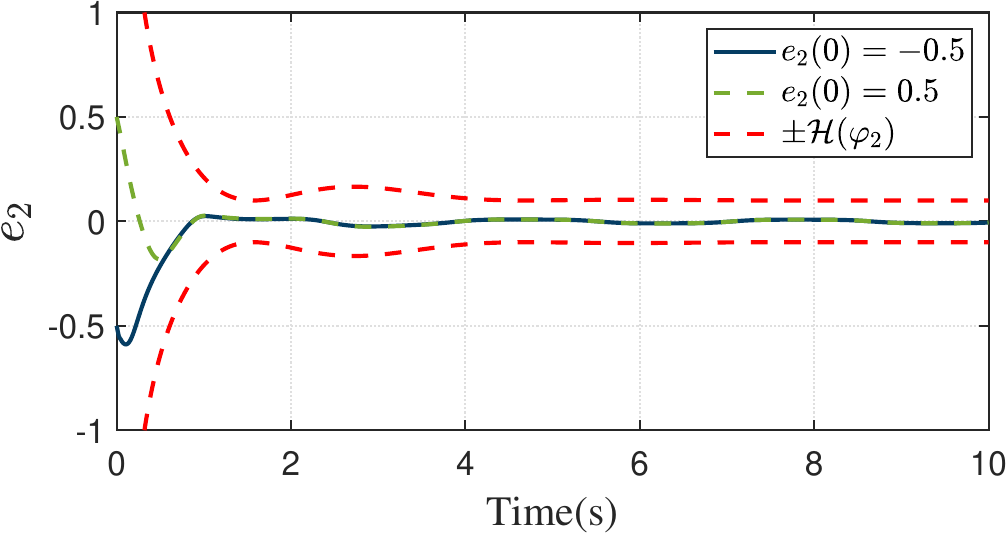}}\\
       {\includegraphics[width=6cm]{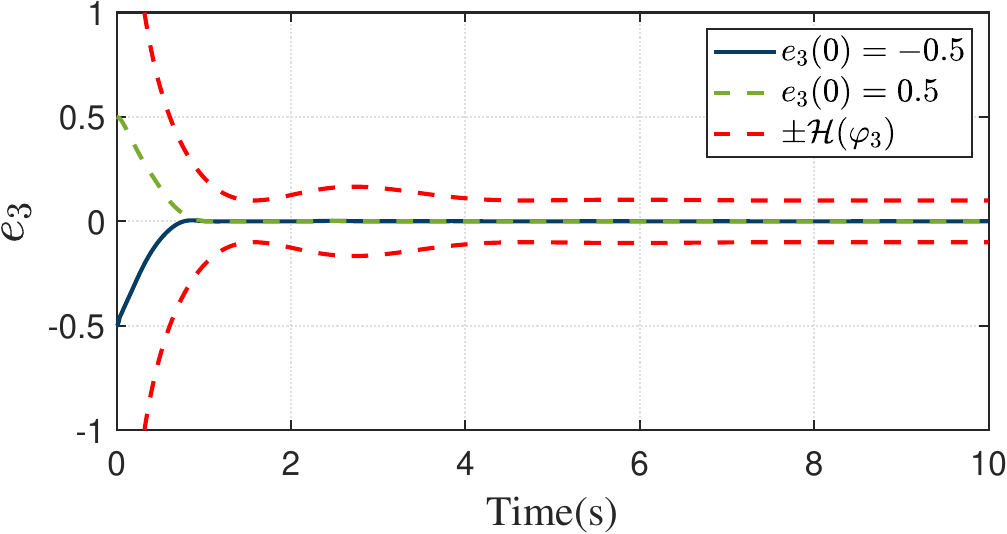}}
 \captionsetup{font={small}}
\caption{The evolution of $e_j$ with different initial conditions, $i=1,2,3$.}
\label{fig-1}
 \end{figure}

 \begin{figure}[htp]
      \centering
       {\includegraphics[width=6cm]{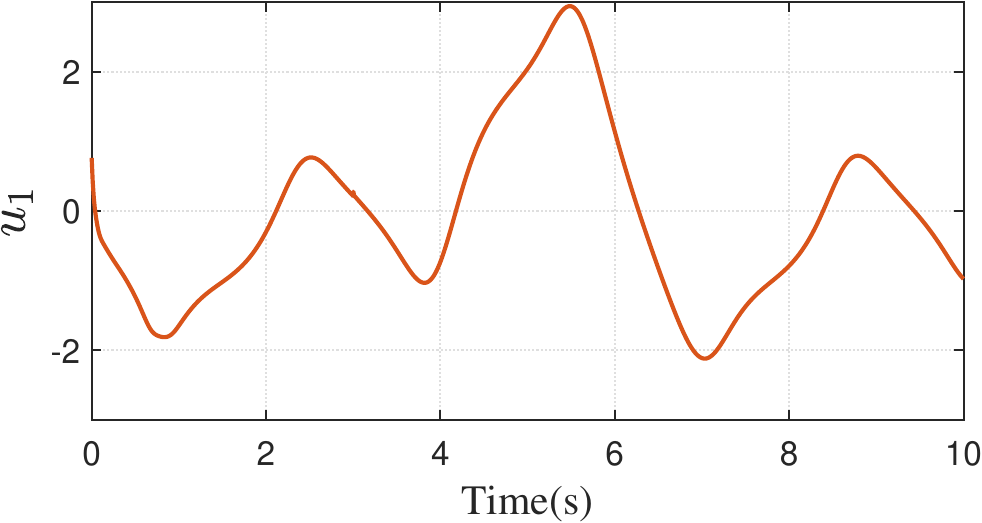}}\\
       {\includegraphics[width=6cm]{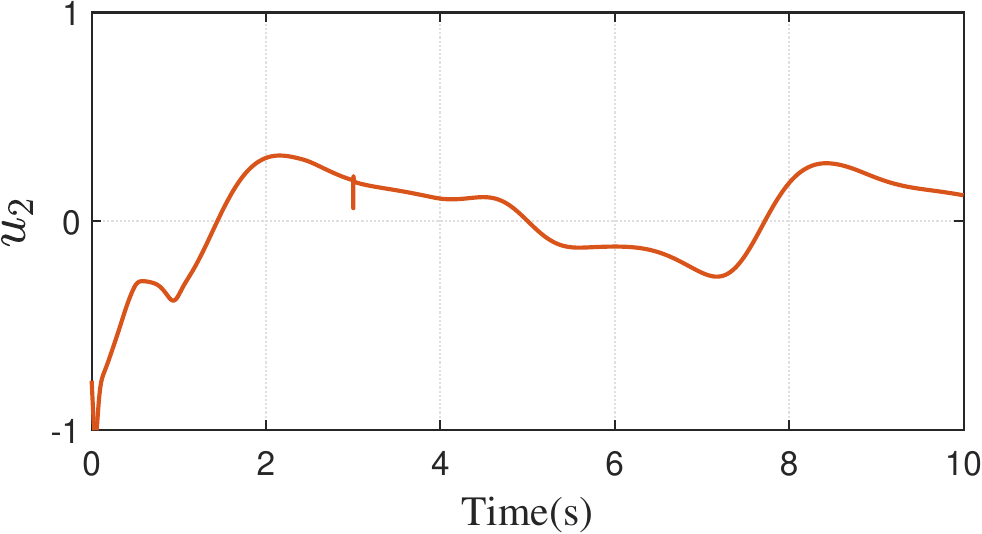}}\\
       {\includegraphics[width=6cm]{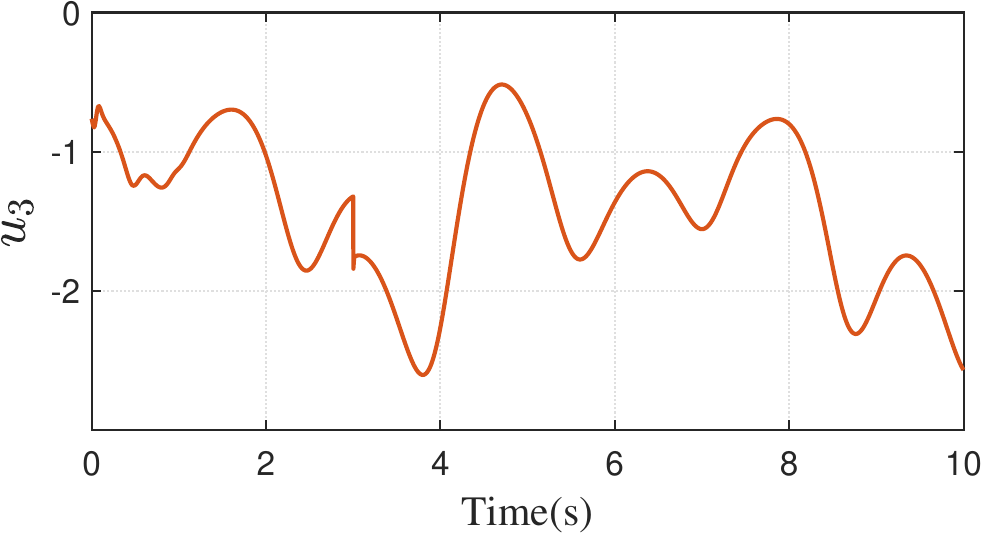}}
 \captionsetup{font={small}}
\caption{Control input signals under $\bm{x}_1(0)=[0.5,0.5,0.5]^T$.}
\label{fig-2}
 \end{figure}
{The rate function is chosen as $\beta(t)=\exp(-0.9t)\cos^2(t)$, and the key design parameters are selected as $\underline{\delta}_j = \bar{\delta}_j=1$,  $\varphi_{j0}=1$, $\varphi_{jf}=0.1$ and $l=0.9936$, $j=1,2,3$. The other parameters are given as: {$\kappa_1=1$},  {$\kappa_2=1$}, $\sigma_1=\sigma_2=0.01$ and $\mu_1=\mu_2=0.1$.
The initial conditions are set as: $\bm{x}_1(0)=[0.5,0.5,0.5]^T$(rad) and $[1.2,-0.5,-0.5]^T$(rad),  $\bm{x}_2(0)=[0,0,0]^T$, $\hat{\theta}_1=\hat{\theta}_2=0$. The simulation results are shown in Figs.\ref{fig-1} and \ref{fig-2}.  Fig.\ref{fig-1} shows that the output tracking errors $e_j$, $j=1,2,3$, evolve within the prescribed performance bounds ($\mathcal{H}(-\varphi_j(t)),\mathcal{H}(\varphi_j(t))$). Fig.\ref{fig-2} shows the boundedness of the control signal $u$.}

\subsection{Application to spacecraft (non-square case)}\label{s52}
In order to validate the effectiveness, we implement the proposed control approach into a spacecraft actuated by four RWs \cite{ShaoX2023auto}, for which the dynamics can be derived as
\begin{equation}\label{sp1}
  \dot{\bm{x}}=\bm{f}(\bm{x})+\bm{g}(\bm{x},t)\bm{u}_a+\bm{d}(t)
\end{equation}
with $\bm{x}=\bm{\omega}$, $\bm{f}(\bm{x})=-\bm{J}^{-1}\bm{\omega}^\times\bm{J}\bm{\omega}$, $\bm{g}(\bm{x},t)=\bm{J}^{-1}\bm{D}$, $\bm{u}_a=\bm{\tau}$ and $\bm{d}(t)=\bm{J}^{-1}\bm{d}_0(t)$, where $\bm{\omega}\in\mathbb{R}^{3}$ is reference angular velocity, $\bm{J}\in\mathbb{R}^{3\times3}$ is the total inertia matrix, $\bm{D}\in\mathbb{R}^{3\times 4}$ (4 is the number of RWs) is the RWs configuration matrix with full-row rank, while $\bm{\tau} \in\mathbb{R}^{4}$ and $\bm{d}_0(t)\in\mathbb{R}^{3}$ denote the control and bounded disturbance torques, respectively. Obviously, this is an non-square system with four control inputs and three outputs.
\begin{figure}[htp]
      \centering
       {\includegraphics[width=6cm]{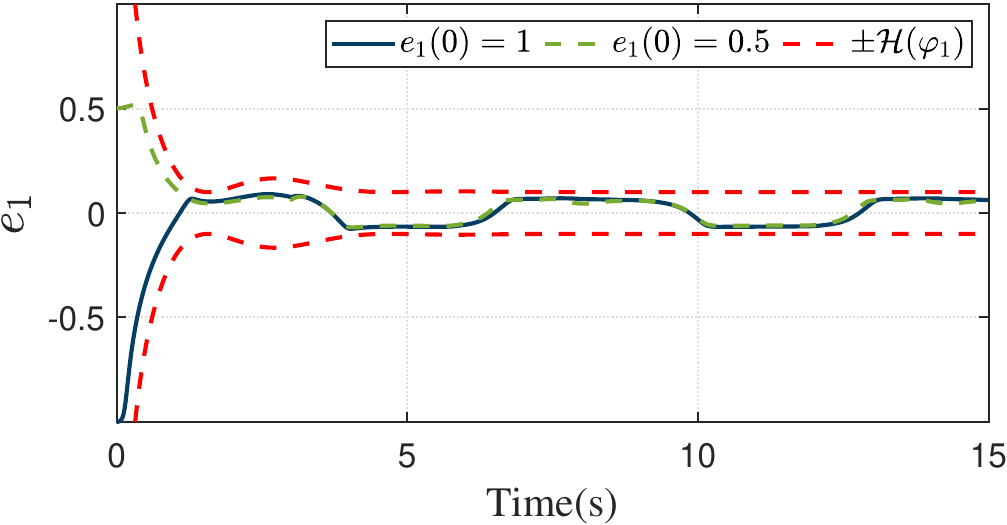}}\\
       {\includegraphics[width=6cm]{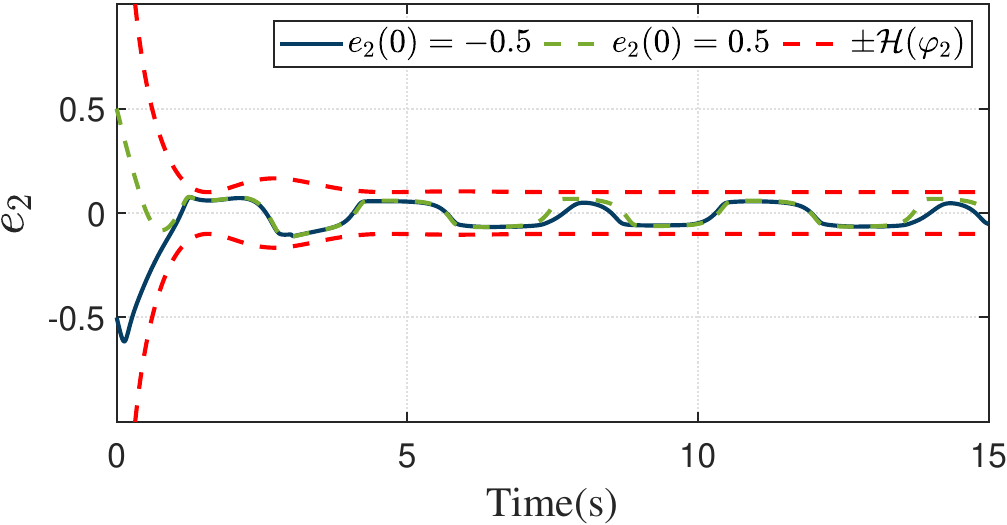}}\\
       {\includegraphics[width=6cm]{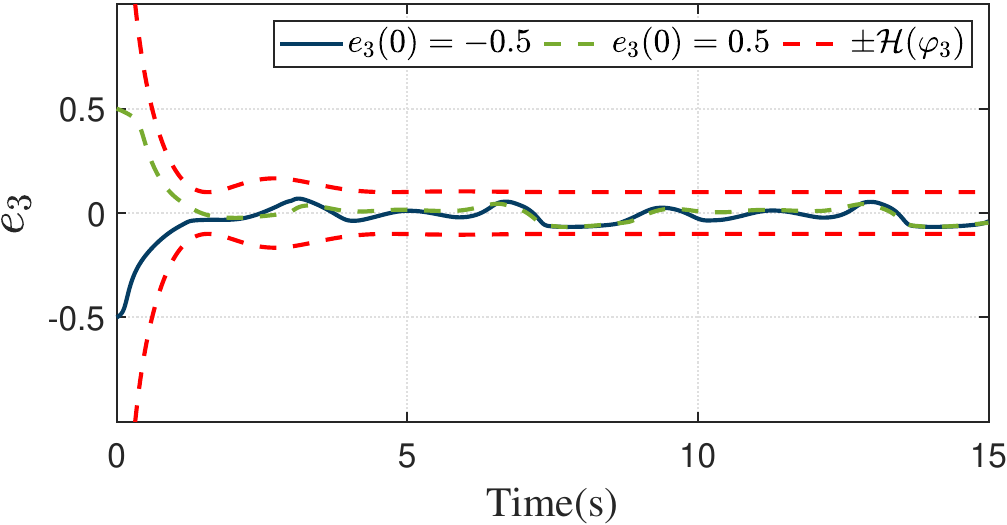}}
 \captionsetup{font={small}}
\caption{The evolution of $e_j$ with different initial conditions, $i=1,2,3$.}
\label{fig-5}
 \end{figure}
\begin{figure}[htp]
  \centering
       {\includegraphics[width=6cm]{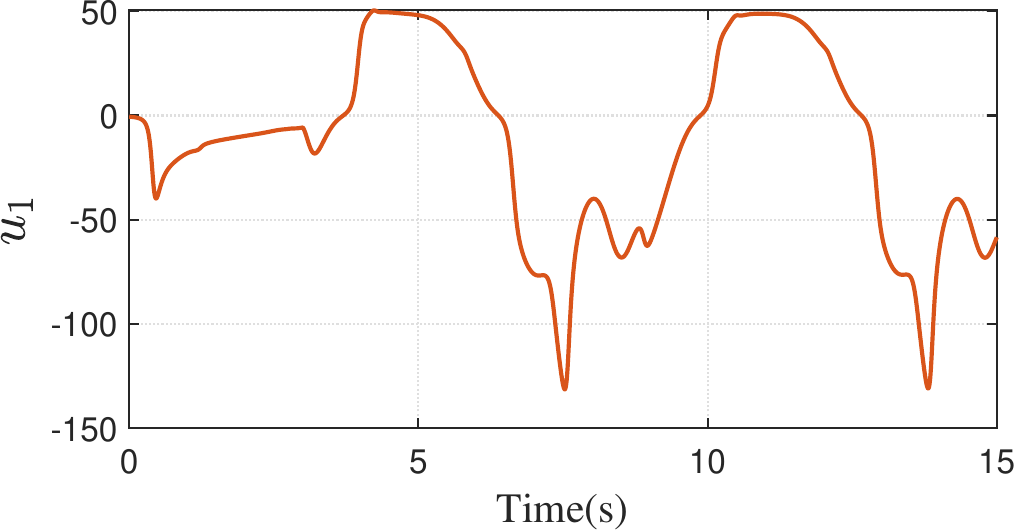}}\\
       {\includegraphics[width=6cm]{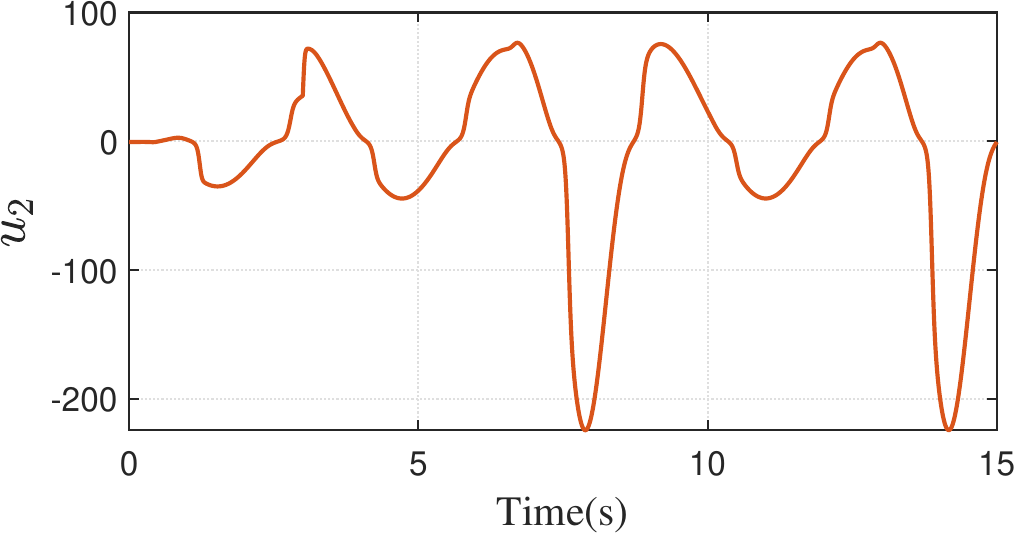}}\\
       {\includegraphics[width=6cm]{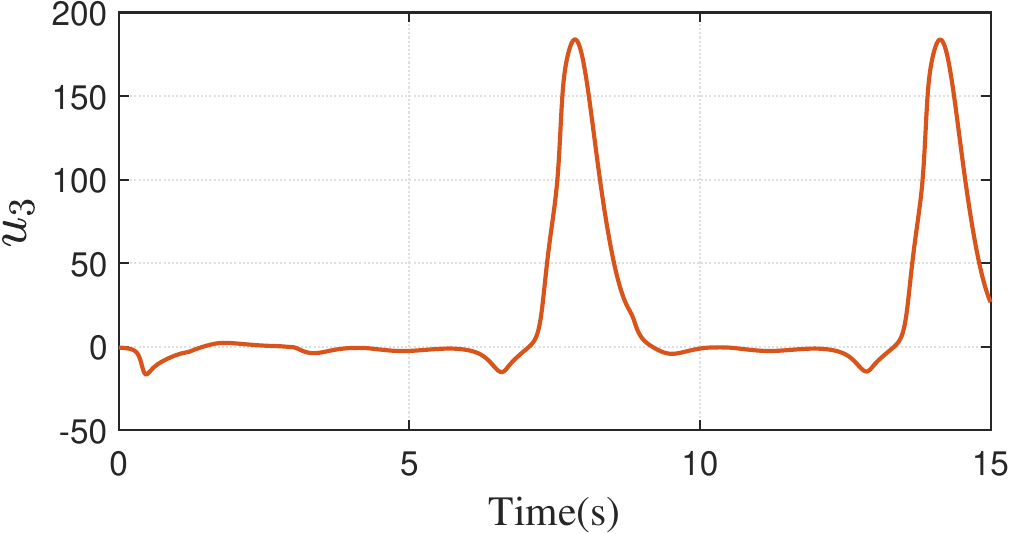}}\\
       {\includegraphics[width=6cm]{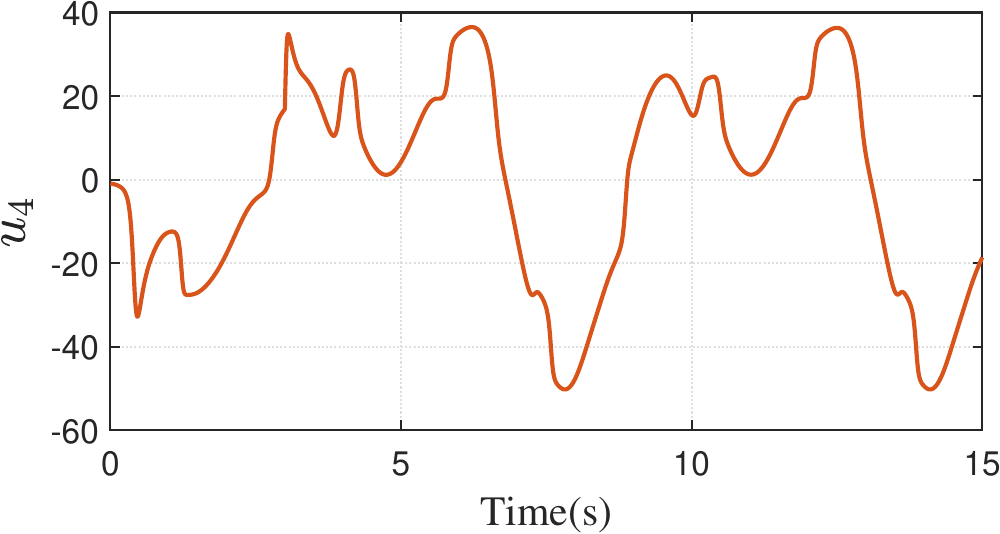}}
 \captionsetup{font={small}}
  \caption{Control input signals under $\bm{x}_1(0)=[1,0.5,0.5]^T$.}
  \label{fig-6}
\end{figure}
The unknown inertia matrix $\bm{J}$ is given by $\bm{J}=\bm{J}_0+\bm{J}_u$, where $\bm{J}_0$ is the nominal part set as
 \begin{align*}
  \bm{J}_0=\begin{bmatrix}
    20 & 1.2 &0.9 \\
    1.2& 5  &1.4\\
    0.9&1.4& 5
  \end{bmatrix}\textrm{kg}\cdot \textrm{m}^2
  \end{align*}
while $\bm{J}_u$  is the uncertain part given by $\bm{J}_u=\textrm{diag}\{0.2e^{-0.2t},\\
2e^{-0.1t},3e^{-0.1t+1}\}\textrm{kg}\cdot \textrm{m}^2$. The configuration matrix is set as
 \begin{equation*}
  \bm{D}=\begin{bmatrix}
    1 & 0 &0&{1}/{\sqrt{3}} \\
    0 & 1 &0&{1}/{\sqrt{3}} \\
    0 & 0 &1&{1}/{\sqrt{3}}
  \end{bmatrix}.
 \end{equation*}
The PLOE fault profiles are outlined as follows:
  \begin{equation*}
  \begin{aligned}
      &\bm{\rho}(t)=\begin{cases}\textmd{diag}\big\{0.5+0.2\sin(t),0.6-0.2\tanh(t),\\
      \ 0.4+0.2\cos(t),0.3-0.1\tanh(t)\big\},\ \ \, t\in(0,3] \\
      \textmd{diag}\big\{0.1-0.08\sin(t),0.2-0.17\sin(t),\\
      0.2-0.15\sin(t),0.2+0.18\sin(t)\big\},\ t\in(3,\infty)\end{cases}\\
     & \bm{\upsilon}(t)= [0.01\tanh(2t),0.01\cos(t),0.01\sin(3t),0.01\cos(2t)]^T.
  \end{aligned}
  \end{equation*}
The key design parameters are selected as $\underline{\delta}_j =\bar{\delta}_j=1$, $j=1,2,3$. The control parameters are given as: {$\kappa_1=1$}, $\sigma_1=0.01$ and $\mu_1=0.1$.  Since configuration matrix $\bm{D}$ can also be interpreted as the actuator allocation matrix, we set $\bm{A}=\bm{D}$. Similarly, it can be verified that ${\bm{{g}}\bm{\rho}\bm{D}^T+\bm{D}\bm{\rho}\bm{{g}}^T}$ is not uniformly positive definite or negative definite for all $t\in(0,\infty]$, but we can still make \emph{Assumption} \ref{A2} hold by choosing $\bm{P}=\textmd{diag}\big\{0.7,0.1,0.6\big\}$.

The rate function is chosen as $\beta(t)=\exp(-0.9t)\cos^2(t)$,  and $\varphi_{j0}=1$, $\varphi_{jf}=0.1$ and $l=0.9936$, $j=1,2,3$.
The initial conditions are set as: $\bm{x}_1(0)=[-0.5,-0.5,-0.5]^T$(rad) and $[1,0.5,0.5]^T$(rad), $\hat{\theta}_1=0$. The simulation results are given in Figs.\ref{fig-5} and \ref{fig-6}.  Fig.\ref{fig-5} shows that the output tracking errors $e_j$, $j=1,2,3$, evolve within the prescribed performance bounds ($\mathcal{H}(-\varphi_j(t)),\mathcal{H}(\varphi_j(t))$). Fig.\ref{fig-6} shows the boundedness of the control signal $u$.

To verify that the proposed control is able to achieve the asymmetric performance behaviors, we select the key design parameters as $\underline{\delta}_j =0.8
,\ \bar{\delta}_j=1$ and $\underline{\delta}_j =1
,\ \bar{\delta}_j=0.6$, $j=1,2,3$. The initial conditions are set as: $\bm{x}_1(0)=[-0.5,-0.5,-0.5]^T$(rad) and $[1,0.5,0.5]^T$(rad). Other parameters and conditions are kept consistent with the previous simulations. The responses of tracking errors are depicted in Fig.\ref{fig-7} and Fig.\ref{fig-8}, which shows that the tracking errors are always evolve within the prescribed performance bounds, in line with the theoretical analysis.
\begin{figure}[htp]
      \centering
       {\includegraphics[width=6cm]{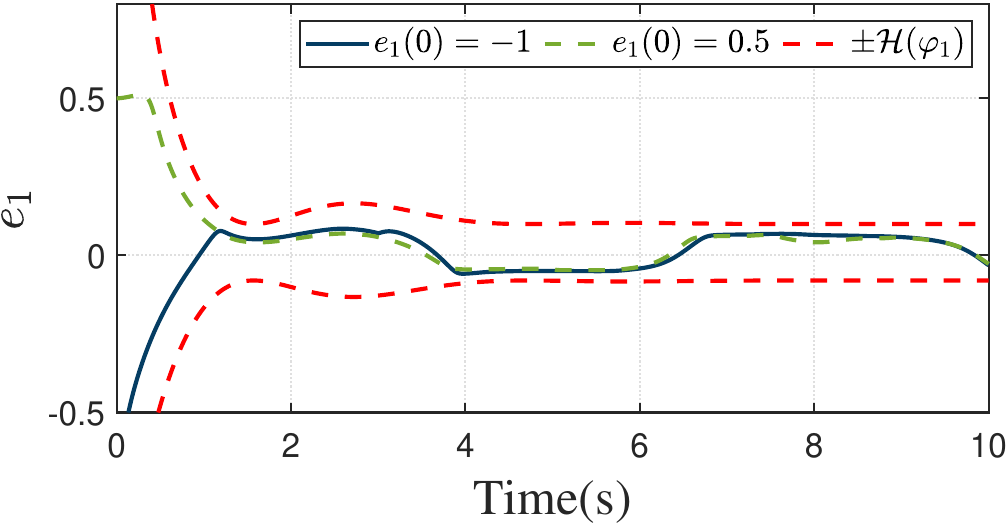}}\\
       {\includegraphics[width=6cm]{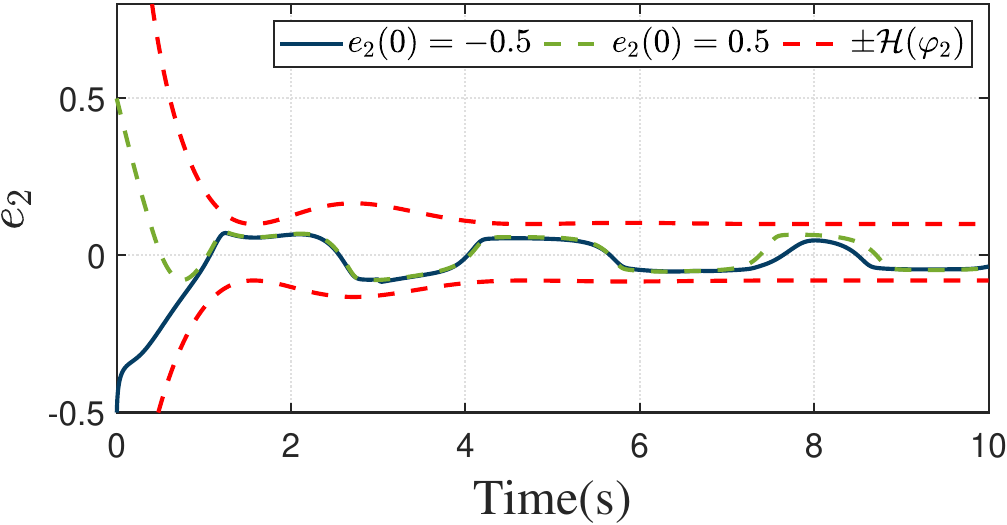}}\\
       {\includegraphics[width=6cm]{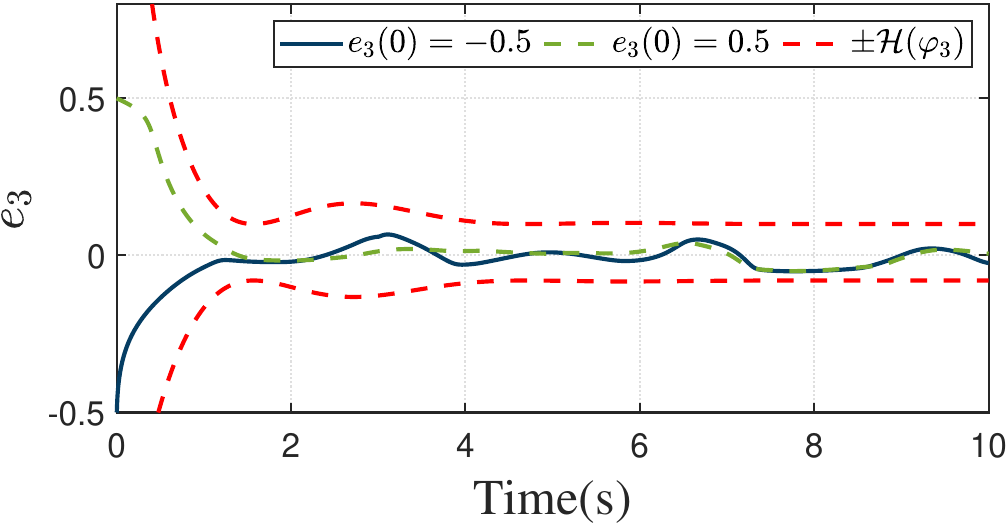}}
 \captionsetup{font={small}}
\caption{The evolution of $e_j$ under $\underline{\delta}_j =0.8
,\ \bar{\delta}_j=1$, $i=1,2,3$.}
\label{fig-7}
 \end{figure}
\begin{figure}[htp]
  \centering
       {\includegraphics[width=6cm]{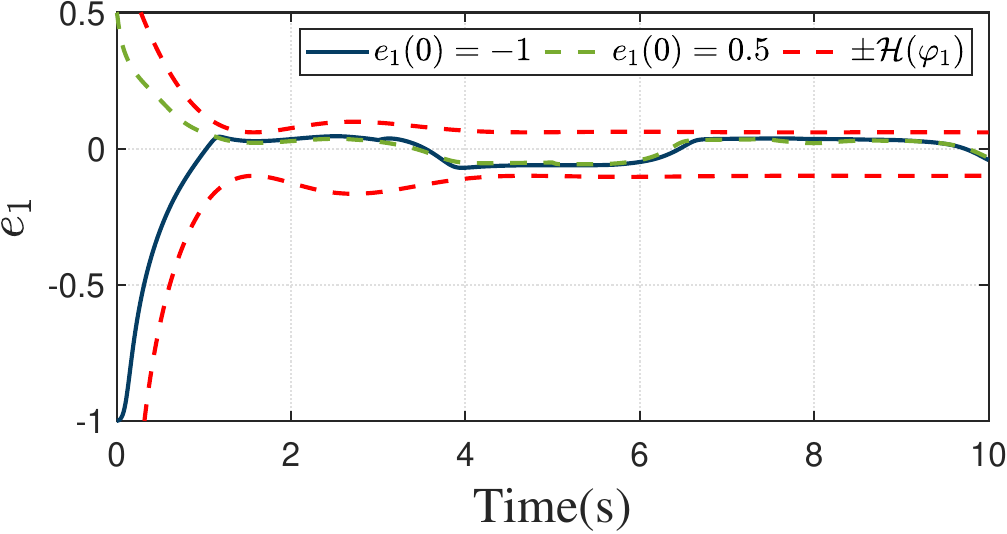}}\\
       {\includegraphics[width=6cm]{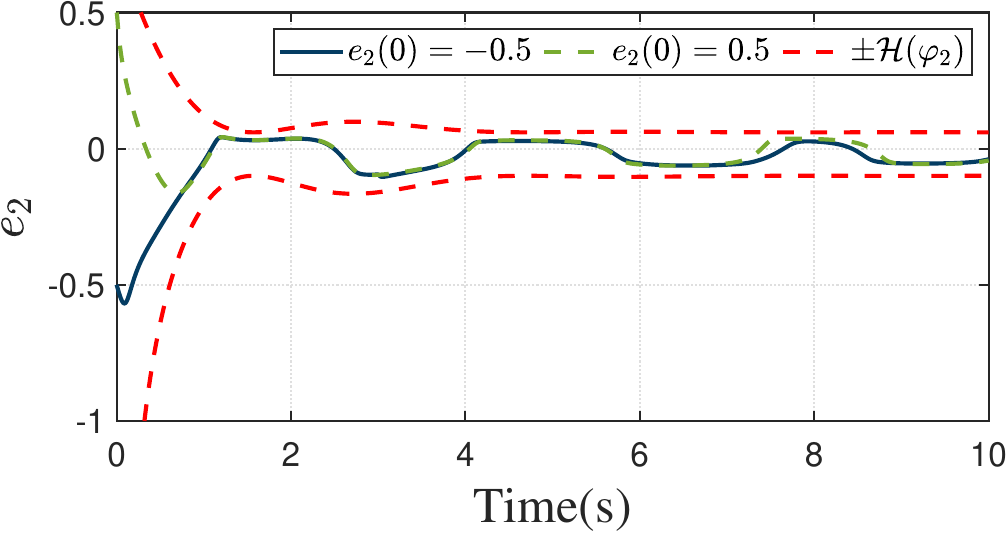}}\\
       {\includegraphics[width=6cm]{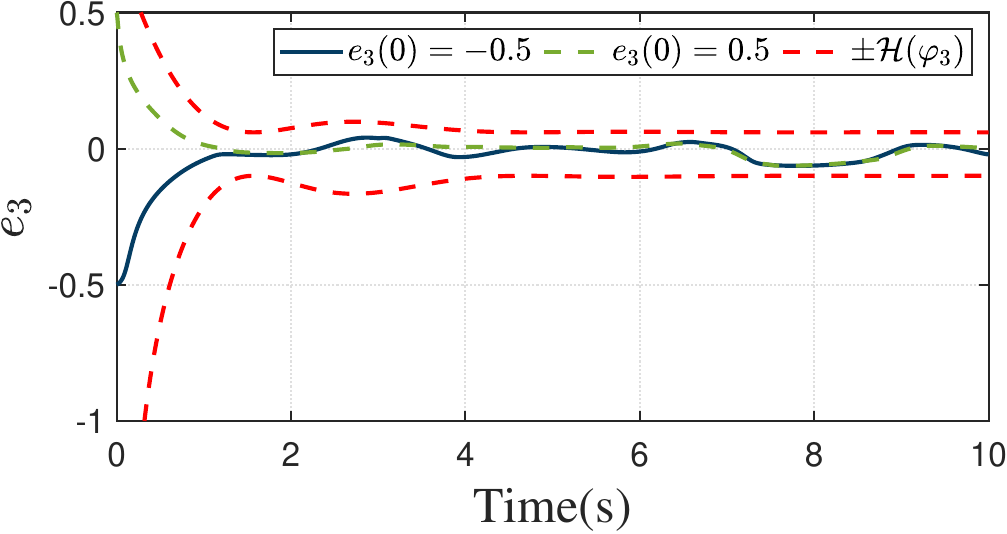}}
 \captionsetup{font={small}}
  \caption{The evolution of $e_j$ under $\underline{\delta}_j =1
,\ \bar{\delta}_j=0.6$, $i=1,2,3$.}
  \label{fig-8}
\end{figure}
\section{Conclusion}
In this paper, a unified prescribed performance tracking control method with generalized controllability condition is proposed for a class of non-square nonlinear strict-feedback systems with actuator faults.
By using a practical matrix decomposition and resorting to some suitable auxiliary matrices, a more relaxed controllability condition for non-square systems with actuator faults is meticulously constructed. The proposed method can achieve global performance and semi-global performance yet asymmetric performance behaviors by choosing key design parameters without changing the control structure.
By constructing some augmented Lyapunov functions using the auxiliary matrices instead of the actuation effectiveness matrix, not only the controllability of the closed-loop system is guaranteed, but also the stringent constraints on the original gain matrix and actuation effectiveness matrix are eliminated.
In the future, we aim to generalize the method to non-square systems in the under-actuated case.
\bibliographystyle{IEEEtran}
\bibliography{reference}

\end{document}